\documentclass[prd,superscriptaddress,amsfonts,amssymb,amsmath,showpacs,twocolumn]{revtex4-2}

\usepackage{bm}
\usepackage{amsfonts}
\usepackage{latexsym}
\usepackage[latin1]{inputenc}
\usepackage{graphicx}
\usepackage{amsmath}
\usepackage{palatino}
\usepackage{mathpazo}
\usepackage{textcomp}
\usepackage{float}
\usepackage{booktabs}
\usepackage{dcolumn}
\usepackage{ragged2e}
\usepackage{hyperref}
\hypersetup{colorlinks,citecolor=blue}
\hypersetup{colorlinks=true,linkcolor=red,filecolor=magenta,    urlcolor=blue}
\usepackage{amsmath}
\usepackage{graphicx}

\usepackage{hyperref}
\usepackage[mathlines]{lineno}
\usepackage{amssymb}
\usepackage{enumitem}
\usepackage{mathtools}
\usepackage{mathrsfs}
\usepackage{setspace}
\usepackage{color}
\usepackage{graphicx}
\usepackage[dvipsnames]{xcolor}
\usepackage{mathalfa}
\usepackage{calligra}
\usepackage[misc]{ifsym}
\DeclareMathAlphabet{\mathpzc}{OT1}{pzc}{m}{it}
\DeclareMathAlphabet{\mathcalligra}{T1}{calligra}{m}{n}
\hypersetup{colorlinks=true,citecolor=blue,linkcolor=red,urlcolor=magenta}
\usepackage{xcolor}
\usepackage{orcidlink}
\usepackage[caption=false]{subfig}
\usepackage{commath}
\def\jnl@style{}
\def\aaref@jnl#1{{\jnl@style#1}}

\def\aaref@jnl#1{{\jnl@style#1}}

\def\aj{\aaref@jnl{AJ}}                   
\def\apj{\aaref@jnl{ApJ}}                 
\def\apjl{\aaref@jnl{ApJ}}                
\def\apjs{\aaref@jnl{ApJS}}               
\def\apss{\aaref@jnl{Ap\&SS}}             
\def\aap{\aaref@jnl{A\&A}}                
\def\aapr{\aaref@jnl{A\&A~Rev.}}          
\def\aaps{\aaref@jnl{A\&AS}}              
\def\mnras{\aaref@jnl{Mon.~Not.~Roy.~Astron.~Soc.}}             
\def\prd{\aaref@jnl{Phys.~Rev.~D}}        
\def\plb{\aaref@jnl{Phys.~Lett.~B}}        
\def\prc{\aaref@jnl{Phys.~Rev.~C}}  
\def\prl{\aaref@jnl{Phys.~Rev.~Lett.}}    
\def\qjras{\aaref@jnl{QJRAS}}             
\def\skytel{\aaref@jnl{S\&T}}             
\def\ssr{\aaref@jnl{Space~Sci.~Rev.}}     
\def\zap{\aaref@jnl{ZAp}}                 
\def\nat{\aaref@jnl{Nature}}              
\def\aplett{\aaref@jnl{Astrophys.~Lett.}} 
\def\apspr{\aaref@jnl{Astrophys.~Space~Phys.~Res.}} 
\def\physrep{\aaref@jnl{Phys.~Rep.}}      
\def\physscr{\aaref@jnl{Phys.~Scr}}       
\def\commat{\aaref@jnl{Comm.~Math.~Phys.}}              
\def\science{\aaref@jnl{Science}}               
\def\cqg{\aaref@jnl{Classical Quant.~Grav.}}            
\def\jpcs{\aaref@jnl{JPCS}}                                     
\def\ijmpd{\aaref@jnl{Int.~J.~Mod.~Phys.~D}}                    
\def\grg{\aaref@jnl{Gen.~Relat.~Gravit.}}               
\def\rpp{\aaref@jnl{Rep.~Prog.~Phys.}}          
\def\npa{\aaref@jnl{Nucl.~Phys.~A}}        
\def\lrr{\aaref@jnl{Living Rev.~Rel.}}                   
\def\jcap{\aaref@jnl{J.~Cosmology Astropart.~Phys.}}    
\def\rmp{\aaref@jnl{Rev.~Mod.~Phys.}}   
\def\epjc{\aaref@jnl{Eur.~Phys.~J.~C}}

\allowdisplaybreaks[1]
\renewcommand{\arraystretch}{1.1}
\begin{document}

\preprint{APS/123-QED}

\title{Geometric structures of Morris-Thorne wormhole metric in $\mathpzc{f}(\mathcal{R},\mathscr{L}_m)$ gravity and energy conditions}

 
\author{V. Venkatesha\orcidlink{0000-0002-2799-2535}}%
 \email{vensmath@gmail.com}
\affiliation{Department of P.G. Studies and Research in Mathematics,
 \\
 Kuvempu University, Shankaraghatta, Shivamogga 577451, Karnataka, INDIA
}%

\author{N. S. Kavya\orcidlink{0000-0001-8561-130X}}
\email{kavya.samak.10@gmail.com}
\affiliation{Department of P.G. Studies and Research in Mathematics,
 \\
 Kuvempu University, Shankaraghatta, Shivamogga 577451, Karnataka, INDIA
}%

\author{P.K. Sahoo\orcidlink{0000-0003-2130-8832}}
\email{pksahoo@hyderabad.bits-pilani.ac.in}
\affiliation{
 Department of Mathematics, Birla Institute of Technology and Science-Pilani,\\
 Hyderabad Campus, Hyderabad 500078, INDIA
}%


\date{\today}

\begin{abstract}
  The aim of this manuscript is to study the traversable wormhole (WH) geometries in the curvature matter coupling gravity. We investigate static spherically symmetric Morris-Thorne WHs within the context of $\mathpzc{f}(\mathcal{R},\mathscr{L}_m)$ gravity. To accomplish this, we examine the WH model in four different cases (i) linear $\mathpzc{f}(\mathcal{R},\mathscr{L}_m)$ model, $\mathpzc{f}(\mathcal{R},\mathscr{L}_m)=\alpha \mathcal{R}+\beta \mathscr{L}_m$ with anisotropic matter distribution having the relation $p_r=m p_t$ (ii) linear $\mathpzc{f}(\mathcal{R},\mathscr{L}_m)$ model having anisotropic matter distribution along with the equation of state parameter, $p_r=\omega \rho$, (iii) non-linear model $\mathpzc{f}(\mathcal{R},\mathscr{L}_m)=\dfrac{1}{2}\mathcal{R}+\mathscr{L}_m^\eta$ with specific form of energy density and (iv) non-linear $\mathpzc{f}(\mathcal{R},\mathscr{L}_m)$ model, $\mathpzc{f}(\mathcal{R},\mathscr{L}_m)=\dfrac{1}{2}\mathcal{R}+(1+\xi \mathcal{R})\mathscr{L}_m$ with isotropic matter distribution and having the linear relation between pressure and energy density, $p=\omega \rho$. Additionally, in the latter case, we consider a specific power-law shape function $b(r)=r_0 \left(\dfrac{r_0}{r}\right)^n$. Furthermore, we analyze the energy conditions for each WH model  to verify their physical viability. As a novel outcome, we can see the validation of the null energy condition for the $\mathpzc{f}(\mathcal{R},\mathscr{L}_m)$ model that suggests ruling out the necessity of exotic matter for the traversability of the WH.  At last, an embedding diagram for each model is illustrated that describes the WH geometry.
\begin{description}
\item[Keywords]
Wormhole, $\mathpzc{f}(\mathcal{R},\mathscr{L}_m)$ gravity, energy conditions,embedding diagram.

\end{description}
\end{abstract}

\maketitle

\section{INTRODUCTION}\label{sectionI}

        \par A wormhole is a passage-like geometric structure that connects different universes or distinct regions of the same universe. In 1916, Flamm proposed the possible existence of such geometry in the manifold \cite{1flamm}. Einstein, with his collaborator Rosen investigated a mathematical alternative to the singularities of spacetime through coordinate transformations \cite{1einsteinrosen}. Here, they replaced Schwarzschild singularity with a bridge-like structure. Morris and Thorne examined mathematical criteria for a traversable wormhole that can make it a time machine \cite{1morrisandthorne}. Further, they investigated that the traversable wormhole violates the null energy conditions. Since ordinary matter satisfies the known laws of physics, violation of NEC indicates the presence of exotic matter \cite{1nec1,1nec2}. However, to achieve a physically viable model, one has to repudiate the existence of the hypothetical fluid. Since it was not possible to rule out the presence of such a candidate in the framework of GR \cite{1exotic1,1exotic2}, an alternative approach was encouraged that could minimize or even nullify the usage of exotic matter \cite{1visser1,1visser2,1kuhfitting}. Recently, it is found that the hypothetical fluid can be effectively dealt with modified theories of gravity. In  the background of $\mathpzc{f}(\mathcal{R})$ gravity, Lobo and Oliveira studied WH structures in $\mathpzc{f}(\mathcal{R})$ \cite{1ref1}. A few prominent results pertaining to the viable WHs can be seen in \cite{1ref2}. Here the discussion is made in the absence of exotic matter. Within the framework of modified gravity theories, Azizi \cite{1azizi}, Rahaman et al \cite{1rahaman}, Zubair et al \cite{1zubair}, \"{O}vg\"{u}n \cite{1ref22a} and Samanta et al \cite{1samanta} have provided significant results. In recent years, the investigation of wormhole solutions is gaining importance. Maldacena and Milekhin in their work constructed humanly traversable WHs using the Randall-Sundram model \cite{1ref7}. Rahaman el al. have explained the possible existence of WHs in the galactic halo region \cite{1frahaman} Within the context of black-bounce spacetime, dynamics of spherically symmetric traversable WHs possessing thin shells are investigated in \cite{1ref8}. In the background of curvature-matter coupling gravity, Garcia and Lobo have studied the behavior of WH in the presence of non-minimal coupling between the Ricci scalar $\mathcal{R}$ and the matter \cite{1ref13}. The studies on WH solutions in teleparallel gravity \cite{1ref15}, WH geometries in extended theories of teleparallel gravity \cite{1ref16,1ref15a}, brane \cite{1ref21}, Einstein-Gauss-Bonnet theory \cite{1ref22}, Brans-Dicke theory \cite{1ref23}, $\mathpzc{f}(\mathcal{R})$ gravity\cite{1ref24}, $\mathpzc{f}(\mathcal{R},\mathcal{T})$ gravity \cite{1ref26}, and Rastall gravity \cite{1ref28} can be seen in the literature.
		\par The $\mathpzc{f}(\mathcal{R})$ theory of gravity can successfully explain many cosmological scenarios \cite{1fofr}. The interpretations of late-time acceleration \cite{1fr1}, ruling-out of DM candidate in analyzing the dynamics of massive test particles \cite{1fr2}, and unification of inflation with DE \cite{1fr3} can be fairly explained using $\mathpzc{f}(\mathcal{R})$ theories. Also, many arguments point to the capability of such higher-order theories, to explain the flatness of galaxies' rotational curves \cite{1fr4}. Motivated by these developments, numerous generalized $\mathpzc{f}(\mathcal{R})$ gravity models came into existence. For instance, Gauss-Bonnet theory, $\mathpzc{f}(\mathcal{R},\mathcal{T})$ theory \cite{1frt},  $\mathpzc{f}(\mathcal{R},\Box \mathcal{R},\mathcal{T})$ theory\cite{1frboxrt}, $\mathpzc{f}(\mathcal{R},\mathcal{T},\mathcal{R}_{\mu\nu}\mathcal{T}^{\mu\nu})$ theory \cite{1frtrmunutmunu} and the like. Among the generalized $\mathpzc{f}(\mathcal{R})$ gravity, the theory that takes into account the influence of an `extra' force by considering the coupling between the matter Lagrangian density and the Ricci Scalar is $\mathpzc{f}(\mathcal{R},\mathscr{L}_m)$ gravity \cite{1frlm}. Interestingly, this extra force, which is orthogonal to the four-velocity, is responsible for the non-geodesic motion of the test particles. Thus, there is a violation of the equivalence principle. With this regard, in \cite{1solarsystem} we can see the constraints on $\mathpzc{f}(\mathcal{R},\mathscr{L}_m)$ gravity models from solar system experiments. In literature, one can see numerous works on curvature-matter coupling gravity theories \cite{1frlm1,1frlm2,1frlm3,1frlm4,1frlm5,1frlm6,1frlm7,1frlm8}.

		\par Motivated by these, we investigate the WH geometries within the realm of $\mathpzc{f}(\mathcal{R},\mathscr{L}_m)$ gravity. Here, we explore the wormhole solution for different $\mathpzc{f}(\mathcal{R},\mathscr{L}_m)$ models such as linear, minimally coupled, and non-minimally coupled models. Further, we interpret the influence of coupling on energy conditions. In sec \ref{sectionII}, we derive the field equations of $\mathpzc{f}(\mathcal{R},\mathscr{L}_m)$ gravity. Sec \ref{sectionIII} gives the criteria for a traversable WH. The field equations of $\mathpzc{f}(\mathcal{R},\mathscr{L}_m)$ gravity for spherically symmetric non-rotating WH is discussed in sec \ref{sectionIV}. Sec \ref{sectionV} presents a generalized linear model of $\mathpzc{f}(\mathcal{R},\mathscr{L}_m)$ gravity. Further, we analyze the non-linear WH model in sec \ref{sectionVI},  and their embedding diagrams are examined in sec \ref{sectionVII}. Finally, the last section \ref{sectionVIII}, presents the overall conclusion and inference of the work.

\section{THE FIELD EQUATIONS IN $\mathpzc{f}(\mathcal{R},\mathscr{L}_m)$ GRAVITY}\label{sectionII}			
			
		\par For $\mathpzc{f}(\mathcal{R},\mathscr{L}_m)$ gravity, the action integral of the gravitational field reads,
		\begin{equation}\label{action}
			S=\int \mathpzc{f}(\mathcal{R},\mathscr{L}_m)	\sqrt{-g}\, d^4x,
		\end{equation}
		where $\mathpzc{f}(\mathcal{R},\mathscr{L}_m)	$ is an arbitrary function of  $\mathcal{R}$ the Ricci scalar, and $\mathscr{L}_m$, the Lagrangian density corresponding to matter and $g$ is the determinant of metric tensor.
		
		\par The variation of the action integral \eqref{action} with respect to the components of the metric tensor $g^{\mu\nu}$ results in obtaining the field equation for $\mathpzc{f}(\mathcal{R},\mathscr{L}_m)$ gravity \cite{1frlm} given by,
		\begin{equation}\label{fieldequation1}
			\begin{split}
				\mathpzc{f}_\mathcal{R}(\mathcal{R},\mathscr{L}_m)\mathcal{R}_{\mu\nu}+(g_{\mu\nu}\nabla_\mu\nabla^{\mu}-\nabla_\mu\nabla_\nu)\mathpzc{f}_\mathcal{R}(\mathcal{R},\mathscr{L}_m)\\-\dfrac{1}{2}\left[\mathpzc{f}(\mathcal{R},\mathscr{L}_m)- \mathpzc{f}_{\mathscr{L}_m}(\mathcal{R},\mathscr{L}_m)\mathscr{L}_m\right]g_{\mu\nu}=\dfrac{1}{2}\mathpzc{f}_{\mathscr{L}_m}(\mathcal{R},\mathscr{L}_m)\mathcal{T}_{\mu\nu}.
			\end{split}
		\end{equation}
		Here, $\mathpzc{f}_{\mathscr{L}_m}(\mathcal{R},\mathscr{L}_m)\equiv\frac{\partial \mathpzc{f}(\mathcal{R},\mathscr{L}_m)}{\partial \mathscr{L}_m}$, $\mathpzc{f}_\mathcal{R}(\mathcal{R},\mathscr{L}_m)\equiv\frac{\partial \mathpzc{f}(\mathcal{R},\mathscr{L}_m)}{\partial \mathcal{R}}$ and $\mathcal{T}_{\mu\nu}$ is the EMT that takes the form,
		\begin{equation}
			\mathcal{T}_{\mu\nu}=-\dfrac{2}{\sqrt{-g}} \dfrac{\delta(\sqrt{-g}\mathscr{L}_m)}{\delta g^{\mu\nu}}.
		\end{equation}  
		\par By considering the explicit form of the gravitational field equation, one can attain an equation for the covariant divergence of EMT as, 
		\begin{equation}
			\nabla^\mu \mathcal{T}_{\mu\nu}=2\left\lbrace \nabla^\mu \text{ln}\left[\mathpzc{f}_{\mathscr{L}_m}(\mathcal{R},\mathscr{L}_m) \right]\right\rbrace \dfrac{\partial \mathscr{L}_m }{\partial g^{\mu\nu}}. 
		\end{equation}
		\par Furthermore, we get,
		\begin{equation}\label{traceoffieldequation}
			\begin{split}
				3\nabla_\mu\nabla^{\mu}\mathpzc{f}_\mathcal{R}(\mathcal{R},\mathscr{L}_m)+\mathpzc{f}_\mathcal{R}(\mathcal{R},\mathscr{L}_m)\mathcal{R}-2\left[\mathpzc{f}(\mathcal{R},\mathscr{L}_m)\right.\\\left. -\mathpzc{f}_{\mathscr{L}_m}(\mathcal{R},\mathscr{L}_m)\mathscr{L}_m\right]=\dfrac{1}{2}\mathpzc{f}_{\mathscr{L}_m}(\mathcal{R},\mathscr{L}_m)\mathcal{T},
			\end{split}
		\end{equation}
		on contracting the field equation \eqref{fieldequation1}. This provides the correspondence between the trace of EMT $\mathcal{T}=\mathcal{T}^\mu_\mu$, matter Lagrangian density $\mathscr{L}_m$ and the Ricci scalar $\mathcal{R}$.
		
		\par Moreover, using the equations \eqref{fieldequation1} and \eqref{traceoffieldequation}, we obtain another form of gravitational field equation for $\mathpzc{f}(\mathcal{R},\mathscr{L}_m)$ gravity, given by
		\begin{equation}\label{fieldquation2}
			\begin{split}
				\mathpzc{f}_\mathcal{R}(\mathcal{R},\mathscr{L}_m)\left( \mathcal{R}_{\mu\nu}-\dfrac{1}{3}\mathcal{R}g_{\mu\nu}\right) + \dfrac{g_{\mu\nu}}{6}\left[\mathpzc{f}(\mathcal{R},\mathscr{L}_m)\right.\\\left. -\mathpzc{f}_{\mathscr{L}_m}(\mathcal{R},\mathscr{L}_m)\mathscr{L}_m\right]=\dfrac{1}{2}\left(\mathcal{T}_{\mu\nu} -\dfrac{1}{3}\mathcal{T}g_{\mu\nu}\right)\mathpzc{f}_{\mathscr{L}_m}(\mathcal{R},\mathscr{L}_m) \\
				+\nabla_\mu\nabla_{\nu}\mathpzc{f}_\mathcal{R}(\mathcal{R},\mathscr{L}_m).
			\end{split}
		\end{equation}	
\section{CRITERIA FOR A TRAVERSABLE WORMHOLE}\label{sectionIII}

		\par According to Morris and Thorne \cite{1morrisandthorne}, in the Schwarzschild coordinates $(t,r,\theta,\phi)$, a spherically symmetric non-rotating WH metric is depicted as, 
		\begin{equation}\label{whmetric}
			ds^2=-e^{2\varphi(r)}dt^2+\dfrac{dr^2}{1-\dfrac{b(r)}{r}  }  + r^2\left(d\theta^2+\text{sin}^2\theta \,d\phi^2\right), 
		\end{equation}  
		where, $\varphi(r)$ and $b(r)$ are respectively the gravitational redshift and shape functions. 
		
		\par For the traversable WH, the event horizon should not exist. In order to achieve this, the value of $\varphi(r)$ should be finite everywhere in the domain. Also, the radial coordinate $r$ takes the values ranging from $r_0($throat radius$>0)$ to $\infty$. The prominence of the redshift function lies in the nature of its derivative with respect to the radial coordinates as it determines the geometrical aspects of WH. 
		
		\par Further, the function $b(r)$ has to satisfy the following criteria for a traversable WH:
		
		\begin{enumerate}[label=$\circ$,leftmargin=*]
			\setlength{\itemsep}{4pt}
			\setlength{\parskip}{4pt}
			\setlength{\parsep}{4pt}
			\item \textit{Throat condition}: The value of the function $b(r)$ at the throat is $r_0$ and hence $1-\frac{b(r)}{r}>0$ for $r>r_0.$ 
			\item \textit{Flaring-out condition:} The radial differential of the shape function, $b'(r)$ at the throat should satisfy, $b'(r_0)<1.$ 
			\item \textit{Asymptotic Flatness condition:} As $r\rightarrow \infty$, $\frac{b(r)}{r}\rightarrow 0$.
		\end{enumerate} 
		\par In addition, for the profound interpretation of traversable WHs, we need to consider $\mathit{\mathit{l}}(r)$, the proper radial distance function, given by, 
		\begin{equation}
			\mathit{\mathit{l}}(r)=\pm \int_{r_0}^r \dfrac{dr}{\sqrt{\dfrac{r-b(r)}{r}}}.
		\end{equation}
		\par Notably, this function must be finite over radial coordinates. Initially, it decreases from the upper universe to the throat. Then, again it increases in magnitude from the throat to the lower universe. In upper and lower universes $l(r)$ attains opposite signs. 	
		
        \section{WORMHOLE SOLUTIONS IN $\mathpzc{f}(\mathcal{R},\mathscr{L}_m)$ GRAVITY}\label{sectionIV}
		
		\par In the present work, for the traversability of WH, we suppose $\varphi(r)$ to be a constant, so that $\varphi'(r)=0$. In other words, we are considering WH with zero tidal force. Further, the matter distribution is presumed to be anisotropic i.e.,
		\begin{equation}\label{energymomentumtensor}
			\mathcal{T}_{\mu\nu}=(\rho+p_t)\mathfrak{u}_\mu \mathfrak{u}_\nu-p_t\,g_{\mu\nu}+(p_r-p_t)\mathfrak{x}_{\mu}\mathfrak{x}_\nu,
		\end{equation}
		where $\rho$, $p_r$ and $p_t$ are respectively the energy density, the radial pressure, and the tangential pressure. Here, $\mathfrak{u}_\mu$ represents a four-velocity vector with unit norm and $\mathfrak{x}_\mu$ represents a space-like unit vector. For the anisotropic case, the radial pressure $p_r$ will be along $u_\mu$ and the tangential pressure will be orthogonal to $\mathfrak{x}_\mu$. Further, we assume that $\rho$, $p_r$, and $p_t$ depend on the radial coordinate $r$.
		
		\par Now, for the WH metric \eqref{whmetric} along with EMT \eqref{energymomentumtensor}, the field equations \eqref{fieldquation2} yield,
		\begin{widetext}
		\begin{eqnarray}
		    \label{fe1}4\mathpzc{f}_\mathcal{R}\dfrac{b'}{r^2}-(\mathpzc{f}-\mathpzc{f}_{\mathscr{L}_m}\mathscr{L}_m)=(2\rho+p_r+2p_t)\mathpzc{f}_{\mathscr{L}_m},\\
		    \label{fe2}	6\mathpzc{f}_\mathcal{R}''\left(1-\dfrac{b}{r} \right)+3\mathpzc{f}_\mathcal{R}'\left(\dfrac{b-rb'}{r^2} \right)+2\mathpzc{f}_\mathcal{R}\left(\dfrac{3b-rb'}{r^3} \right) 
				-(\mathpzc{f}-\mathpzc{f}_{\mathscr{L}_m}\mathscr{L}_m)=(-\rho-2p_r+2p_t)\mathpzc{f}_{\mathscr{L}_m}, \\
			\label{fe3}6\dfrac{\mathpzc{f}_\mathcal{R}''}{r}\left(1-\dfrac{b}{r} \right)-\mathpzc{f}_\mathcal{R}\left(\dfrac{3b-rb'}{r^3} \right)-(\mathpzc{f}-\mathpzc{f}_{\mathscr{L}_m}\mathscr{L}_m)=(-\rho+p_r-p_t)\mathpzc{f}_{\mathscr{L}_m} ,
		\end{eqnarray}
		\end{widetext}
		
		where primes represent the derivatives with respect to the radial component $r$. Here we take matter Lagrangian $\mathscr{L}_m$ as a function of energy density, 
		\begin{equation}\label{matterlagrangian}
			\mathscr{L}_m(\rho)=\rho(r).
		\end{equation}
        The above choice of $\mathscr{L}_m$ is considered in \cite{lm1}. One can also consider any other form of matter Lagrangian, say $\mathscr{L}_m\equiv\mathscr{L}_m(p)$ or $\mathscr{L}_m\equiv\mathscr{L}_m(\rho,p)$. However, $\mathscr{L}_m\equiv\mathscr{L}_m(\rho)$ is more appropriate choice for $\mathscr{L}_m$. Particularly, we take, $\mathscr{L}_m=\rho$. In literature, many studies have been presented based on the choice of $\mathscr{L}_m$ \cite{lm2,lm3,lm4}. 
		
		\subsection{Energy Conditions}
		
		\par  ECs interpret the physical phenomenon of motion of energy and matter that arise as a result of the Raychaudhuri equation. In \cite{1frlm4} authors have studied ECs in $\mathpzc{f}(\mathcal{R},\mathscr{L}_m)$ gravity. To examine the geodesic motion, we shall consider the criterion for different ECs. For the anisotropic matter distribution \eqref{energymomentumtensor} with $\rho$, $p_r$ and $p_t$ being energy density, radial pressure and tangential pressure, then we have the following:
		\begin{enumerate}[label=$\circ$,leftmargin=*]
			\setlength{\itemsep}{4pt}
			\setlength{\parskip}{4pt}
			\setlength{\parsep}{4pt}
			\item \textit{Null Energy Conditions (NECs)}: Both $\rho+p_t$ and $\rho+p_r$ are non negative.
			\item \textit{Weak Energy Conditions (WECs)}: For non negative energy density, it implies $\rho+p_t$ and $\rho+p_r$ are both non negative.
			\item \textit{Strong Energy Conditions (SECs)}: For non negative $\rho+p_j$,  $\rho+\sum_j p_j$ is non negative $   \ \forall\ j$.
			\item \textit{Dominant Energy Conditions (DECs)}: For non negative energy density, it implies $\rho-|p_r|$ and $\rho-|p_t|$ are both non negative.
		\end{enumerate}
		
\section{Linear model: Revisiting GR}\label{sectionV}
		\par Analogous to GR, we consider a linear model of  $\mathpzc{f}(\mathcal{R},\mathscr{L}_m)$ gravity given by
		\begin{equation}\label{linearmodel}
			\mathpzc{f}(\mathcal{R},\mathscr{L}_m)=\alpha \mathcal{R}+\beta \mathscr{L}_m,
		\end{equation}
		where $\alpha$ and $\beta$ are arbitrary scalar constants. As this model closely resembles the GR scenario, it is quite remarkable. But it reduces to GR only if $\alpha=(1/2)n$ and $\beta = n$, where $n$ is any integer. This particular scenario is studied in literature as a GR case. However, values of the model parameters other than those specified above represent the linear $f(R, L_m)$ model. In our paper, we investigated the case when model parameters could have any real value.  Taking into account the above-defined function, the field equation \eqref{fe1}-\eqref{fe3} yields,
		\begin{align}
			\label{linear_fe1}\dfrac{2\alpha b'}{r^2} &=\beta (2\rho+p_r+2p_t),\\
			\label{linear_fe2}\dfrac{2\alpha(2b'r-3b)}{r^3} &=\beta(\rho+2p_r-2p_t),\\
			\label{linear_fe3}\dfrac{\alpha(rb'+3b)}{r^3} &=\beta(\rho-p_r+p_t).
		\end{align}
		\par Solving these equations we get,
		\begin{align}
			\label{rho}\rho &= \dfrac{2\alpha b'}{\beta r^2},\\
			\label{pr}p_r &= -\dfrac{2\alpha b}{\beta r^3},\\ 
			\label{pt}p_t &= \alpha\dfrac{- r b'+b}{\beta r^3}.
		\end{align}
		\par 
	
 \begin{figure}[t!]
		    \centering
		    \includegraphics[width=0.8\linewidth]{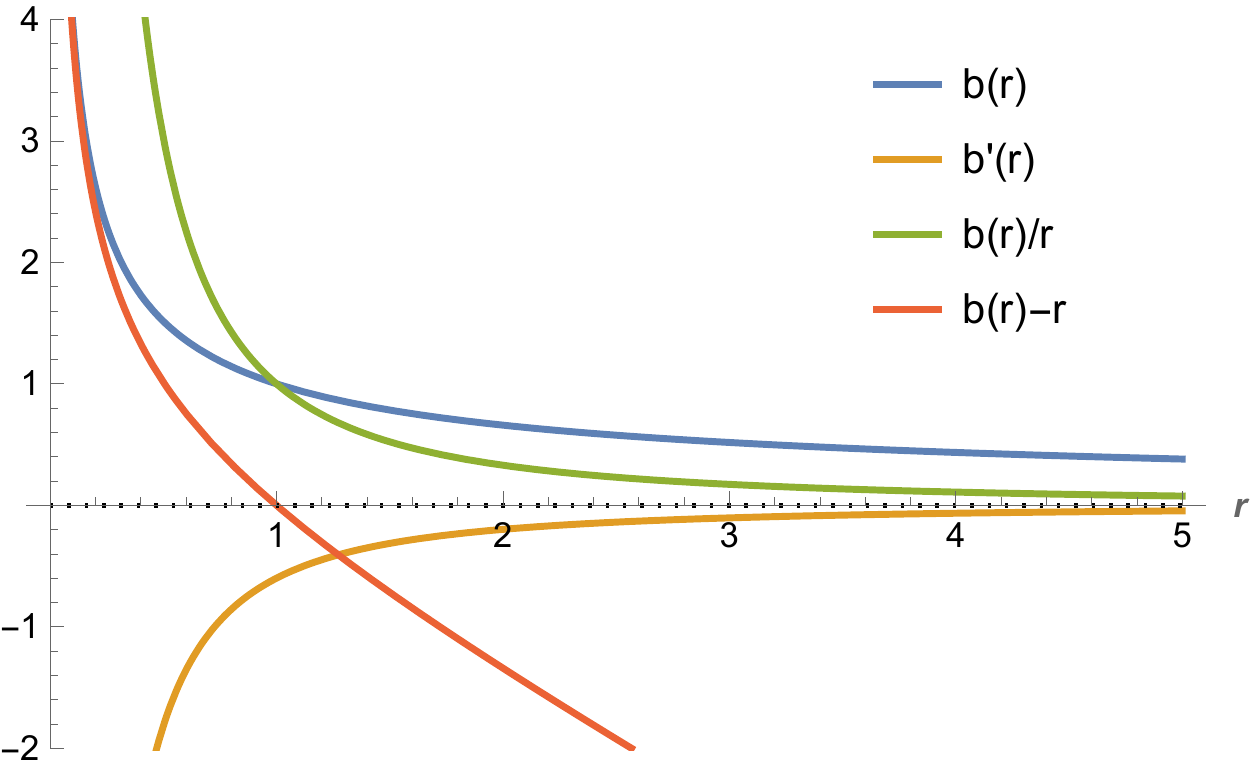}
		    \caption{WHM1: Profile of the shape function $b(r)$ with $m =-1.25$, $k_1=1$ and satisfying $\frac{b(r)}{r}<1$, $b'(r)<1$ and $\frac{b(r)}{r}\rightarrow 0$ as $r\rightarrow \infty$}
		    \label{fig:sfA}
		\end{figure}

 \begin{figure}[H]
		    \centering
		    \includegraphics[width=0.7\linewidth]{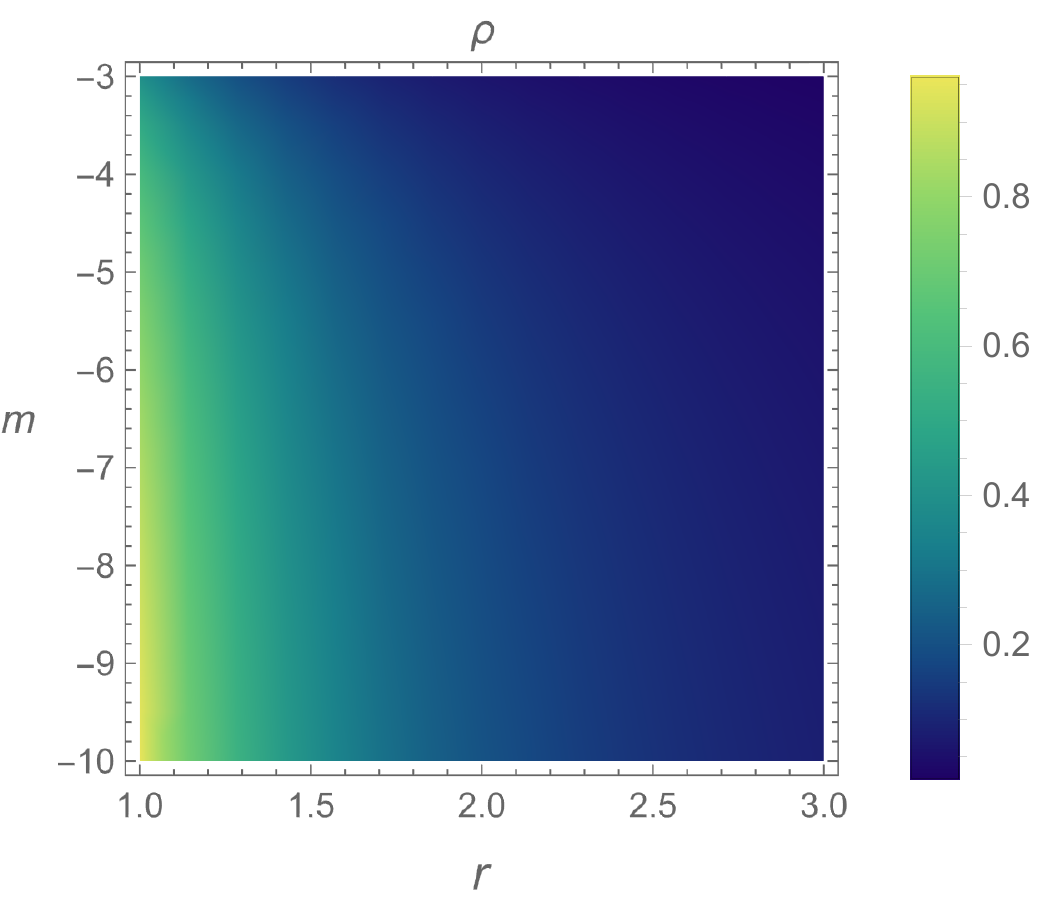}
		    \caption{WHM1: Profile of energy density with $\alpha=3, \beta=5$ and $k_1=1$.}
		    \label{fig:Arho}
		\end{figure}
  \begin{table}[h]
		\caption{The interpretation of energy conditions for case A with the set of values for $m$, $\alpha$, and $\beta$ for which $\rho>0$. In all intervals $\rho+p_r+2p_t=0$.}
		    \label{tab:table1}
		\begin{ruledtabular}
		    \centering
		    \begin{tabular}{c c c c}
				 $m $ & $(-\infty,-3]$ & $(-3,-2)$ & $(-2,-1)$\\ 
				\hline
				$\alpha$ and $\beta$ & $\alpha,\beta>0$ or  & $\alpha,\beta>0$ or  & $\alpha>0,\beta<0$ or  \\
				  &  $\alpha,\beta<0$  &  $\alpha,\beta<0$ & $\alpha<0,\beta>0$ \\
				\hline
				$\rho+p_r$ & violated & violated & obeyed\\
				$\rho+p_t$ & obeyed  & obeyed & violated\\
				$\rho-\mid p_r\mid$ & violated  & violated & violated\\
				$\rho-\mid p_t\mid$ & obeyed  & violated & violated\\
			\end{tabular}
		   \end{ruledtabular}
		\end{table}
  
 \subsection{Case: $p_r=m  p_t$ (WHM1)}	
         In this section, we consider a case in which the radial pressure $p_r$ varies proportionally to the tangential pressure $p_t.$ This is given by,
		\begin{equation}\label{model1}
			p_r=m  p_t,
		\end{equation}
		where `$m $' is a constant. Solving \eqref{pr} and \eqref{pt} in view of \eqref{model1} yields an expression for the shape function $b(r)$ that takes the form,
		\begin{equation}\label{shapefunctionmainmodel1}
			b(r)=k_1\, r^{\frac{2+m }{m }},
		\end{equation}
		with $k_1$ being the constant of integration. Clearly, for $m<0$, one can verify that $b(r)$ satisfies all the necessary criteria for a traversable WH i.e., $\frac{b(r)}{r}<1$, $b'(r)<1$ and $\frac{b(r)}{r}\rightarrow 0$ as $r\rightarrow \infty$. For $m=-1.25$ and $k_1=1$ the profile of the shape function is illustrated in \figureautorefname~ \ref{fig:sfA}. Here, $b(r)-r=0$ for $r=1$ implying, the throat radius $r=r_0=1$.

		\par Now, substituting the newly obtained shape function \eqref{shapefunctionmainmodel1} in equations \eqref{rho}-\eqref{pt}, we get,
		\begin{align}
			\rho &= 2\left(\dfrac{(m +2)\alpha k_1\; r^{-2+2/m }}{\beta m } \right),\\
			p_r &= -\dfrac{2\alpha k_1\: r^{-2+2/m }}{\beta},\\
			p_t &= -\dfrac{2\alpha k_1\: r^{-2+2/m }}{\beta m }.
		\end{align}
        
		\par
\begin{widetext}
		  \begin{figure*}[t!]
	    \centering 
	    \includegraphics[width=0.35\linewidth]{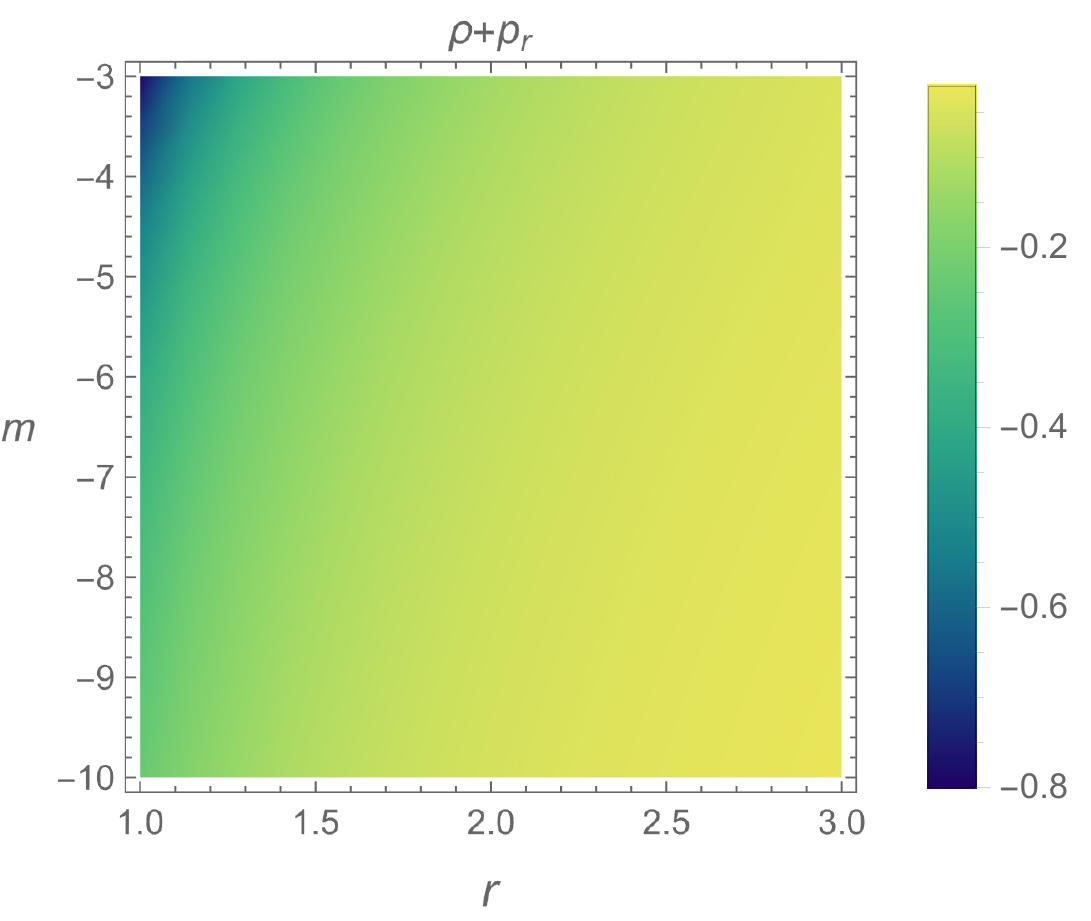}
	    \includegraphics[width=0.35\linewidth]{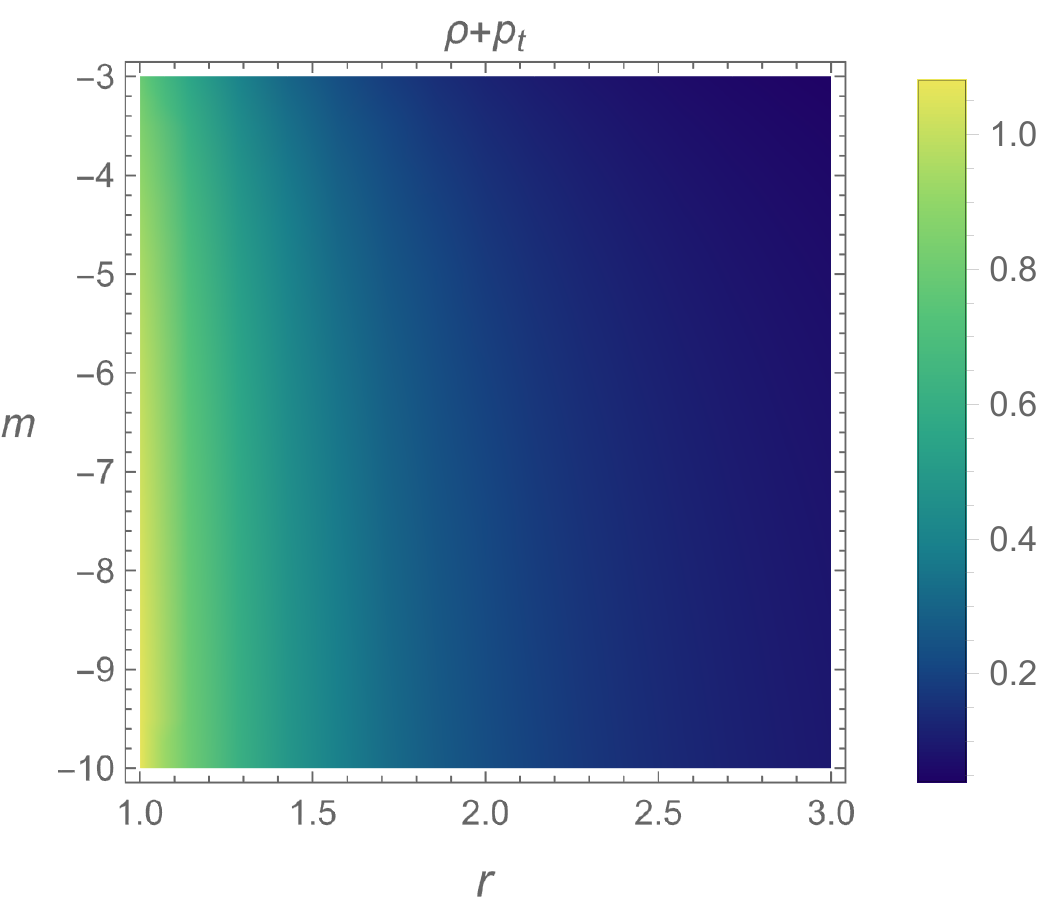}\\
	    \includegraphics[width=0.35\linewidth]{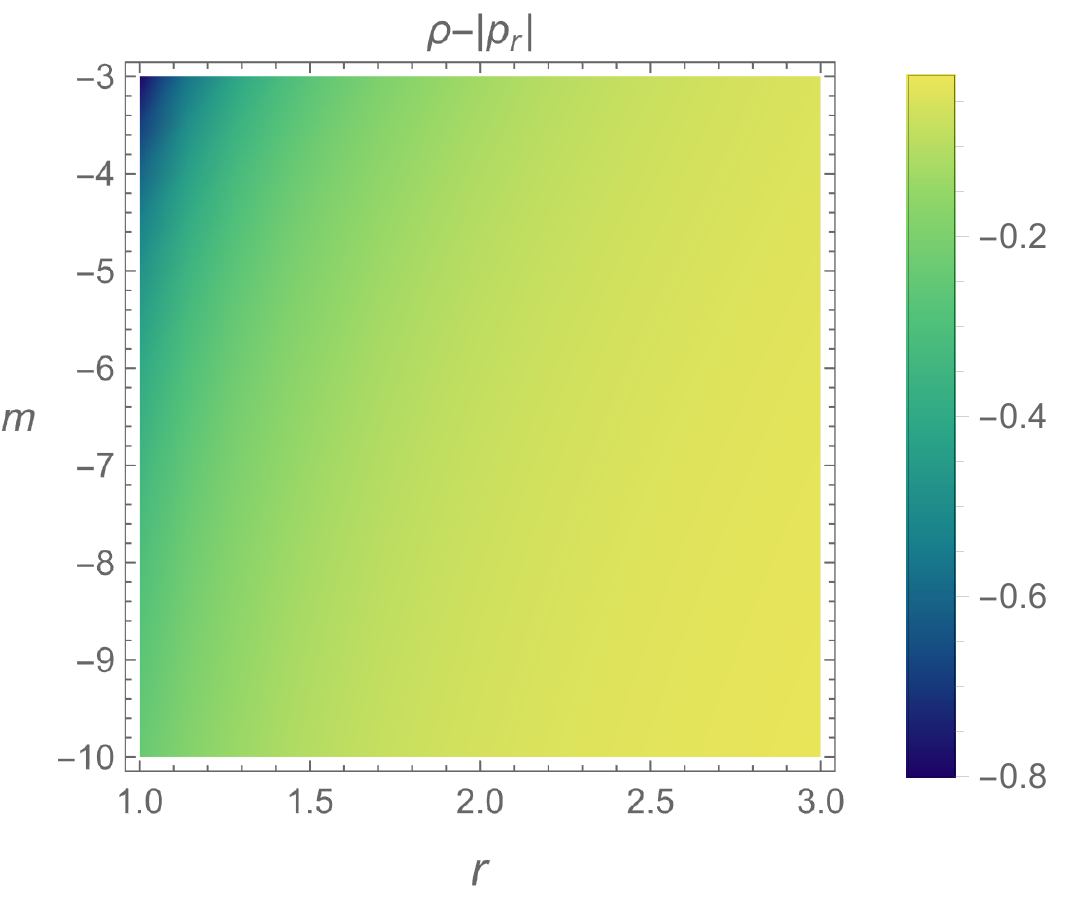}
	    \includegraphics[width=0.35\linewidth]{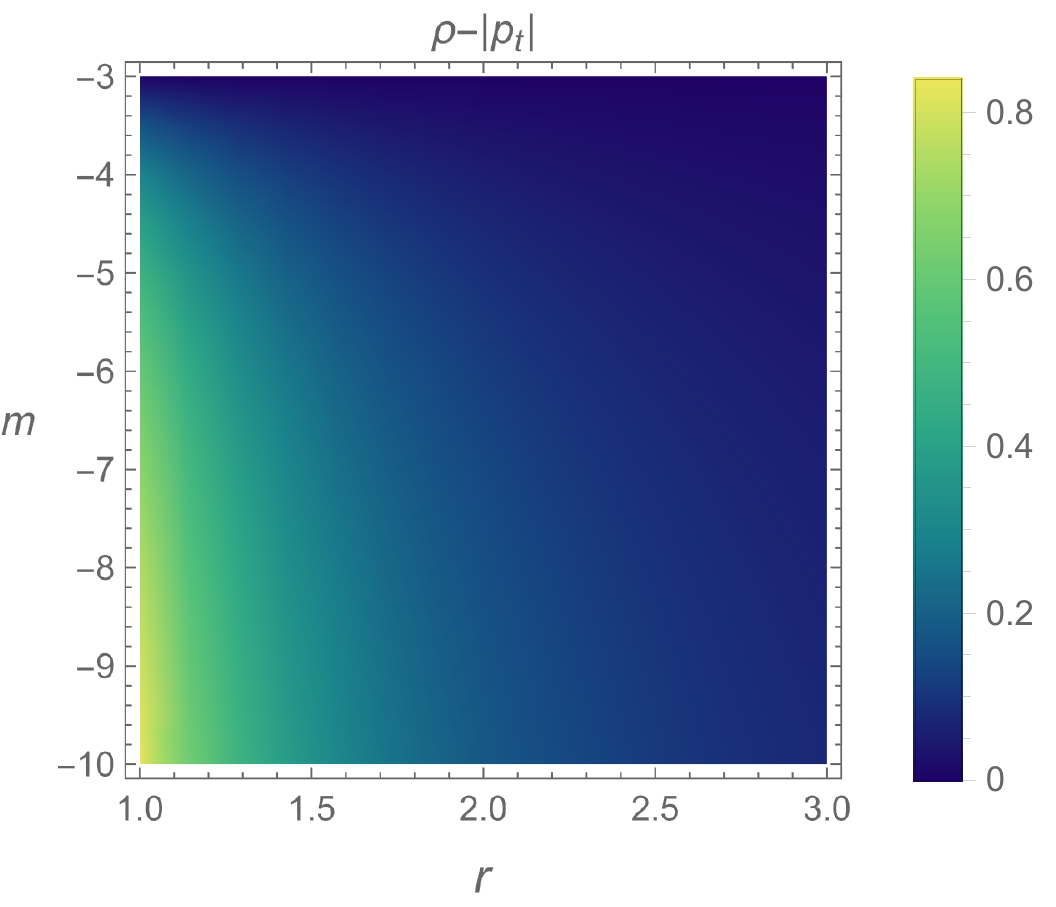}
	    \caption{WHM1: Profile of ECs with $\alpha=3, \beta=5, k_1=1$}
	    \label{fig:Aec}
	\end{figure*}

		\end{widetext}

  We take only those intervals of model parameters $m$, $\alpha$, and $\beta$ for which energy density remains positive. \tableautorefname~ \ref{tab:table1} represents the tabulated result for the NEC and DEC. Also, it is to be noted that, $\rho+p_r+2p_t=0$ in the entire domain. These ECs are depicted in the \figureautorefname~ \ref{fig:Aec} for the values $\alpha=3, \beta=5$ and $k_1=1$. 
	\subsection{Case: $p_r=\omega \rho$ (WHM2)}	
        In this section, we shall presume that the radial pressure $p_r$ varies linearly with respect to the energy density $\rho$. This is related by,
		\begin{equation}\label{model2}
			p_r=\omega \rho,
		\end{equation}
		where `$\omega$' represents the equation of state parameter. Solving \eqref{pr} and \eqref{pt} in view of \eqref{model2} yields an expression for the shape function $b(r)$ that reads,
		\begin{equation}\label{shapefunctionmainmodel2}
			b(r)=k_2\, r^{-\frac{1}{\omega}},
		\end{equation}
		where $k_2$ is a constant of integration. 
		
		\par
\begin{figure}[h]
		    \centering
		    \includegraphics[width=0.8\linewidth]{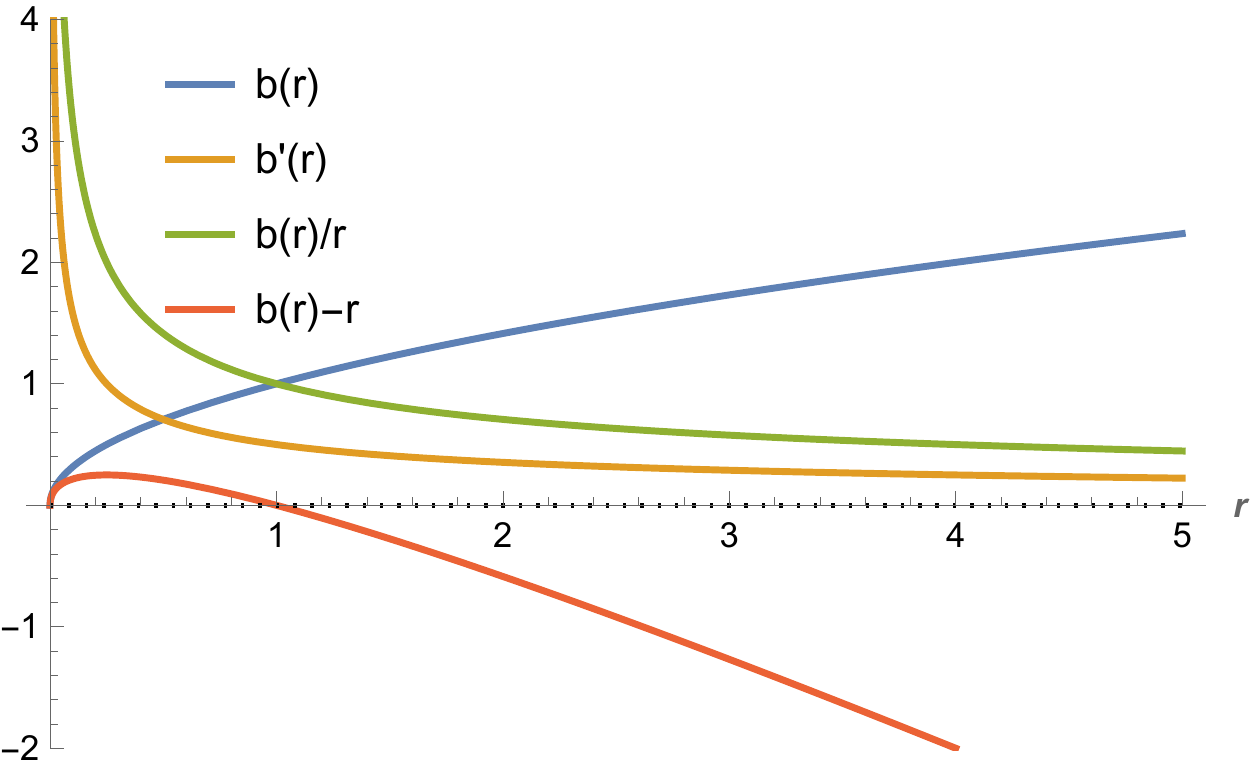}
		    \caption{WHM2: Profile of the shape function $b(r)$ with $\omega =-2$, $k_2=1$ and satisfying $\frac{b(r)}{r}<1$, $b'(r)<1$ and $\frac{b(r)}{r}\rightarrow 0$ as $r\rightarrow \infty$.}
		    \label{fig:sfB}
		\end{figure}

 Here, the WH is traversable if $\omega$ takes the value in the range $\mathbb{R}\backslash[-1,0]$. In particular, the plausible value of the equation of state parameter is, $\omega \in (-\infty,-1)\cup(0,1]$. For $\omega =-2$ and $k_2=1$ the profile for the shape function with satisfying conditions is illustrated in the \figureautorefname~ \ref{fig:sfB}.
			
        \begin{figure}[h]
		    \centering
		    \includegraphics[width=0.7\linewidth]{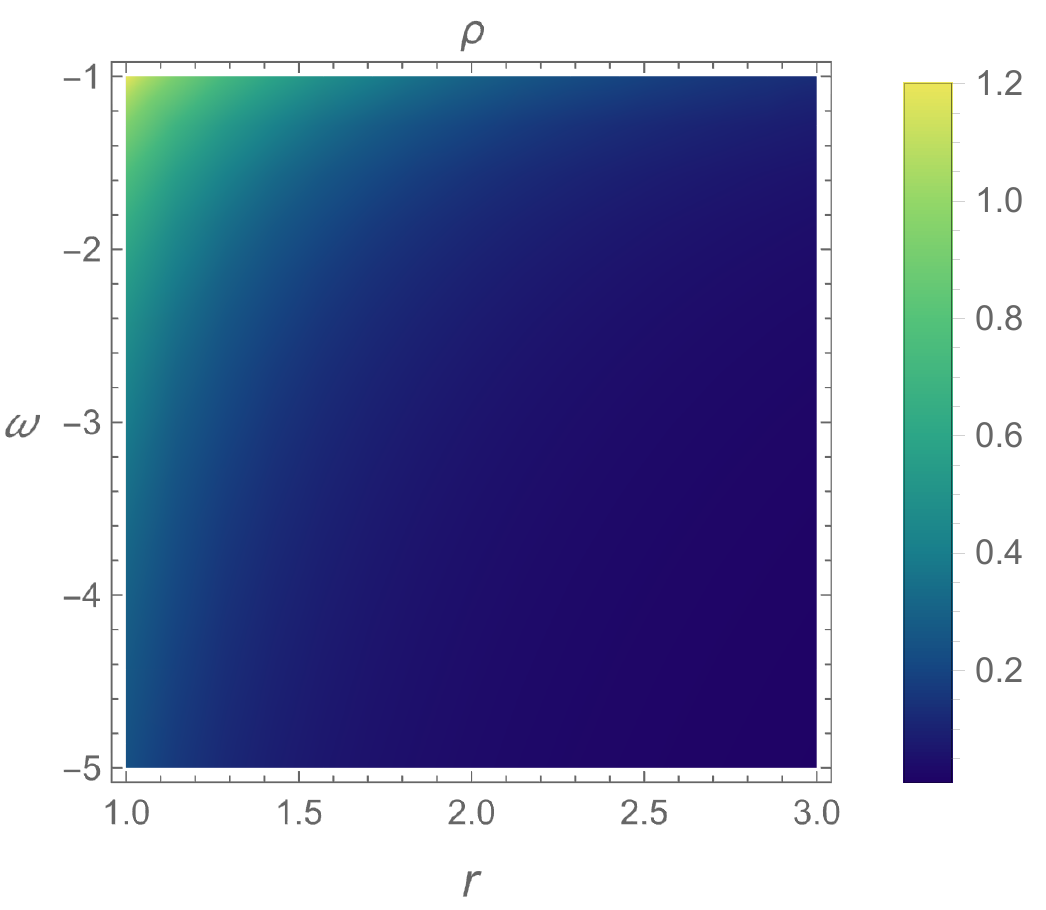}
		    \caption{WHM2: Profile of energy density with $\alpha=3, \beta=5, k_2=1$}
		    \label{fig:Brho}
		\end{figure}
		
		\par Substituting the expression \eqref{shapefunctionmainmodel2} for shape function in corresponding equations of physical parameters \eqref{fe1}-\eqref{fe3} one can get,
		\begin{align}
			\label{rhomodel2}\rho &= \frac{-2 k_{2} \alpha  r^{-3-\frac{1}{\omega}}}{\beta \omega},\\
			\label{prmodel2}p_r &= \frac{-2 k_{2} \alpha  r^{-3-\frac{1}{\omega}}}{\beta },\\
			\label{ptmodel2}p_t &= \frac{k_{2}  \alpha (\omega+1) r^{-3-\frac{1}{\omega}}}{\beta \omega } .
		\end{align}
		\par
\begin{figure*}[t!]
	    \centering 
	    \includegraphics[width=0.35\linewidth]{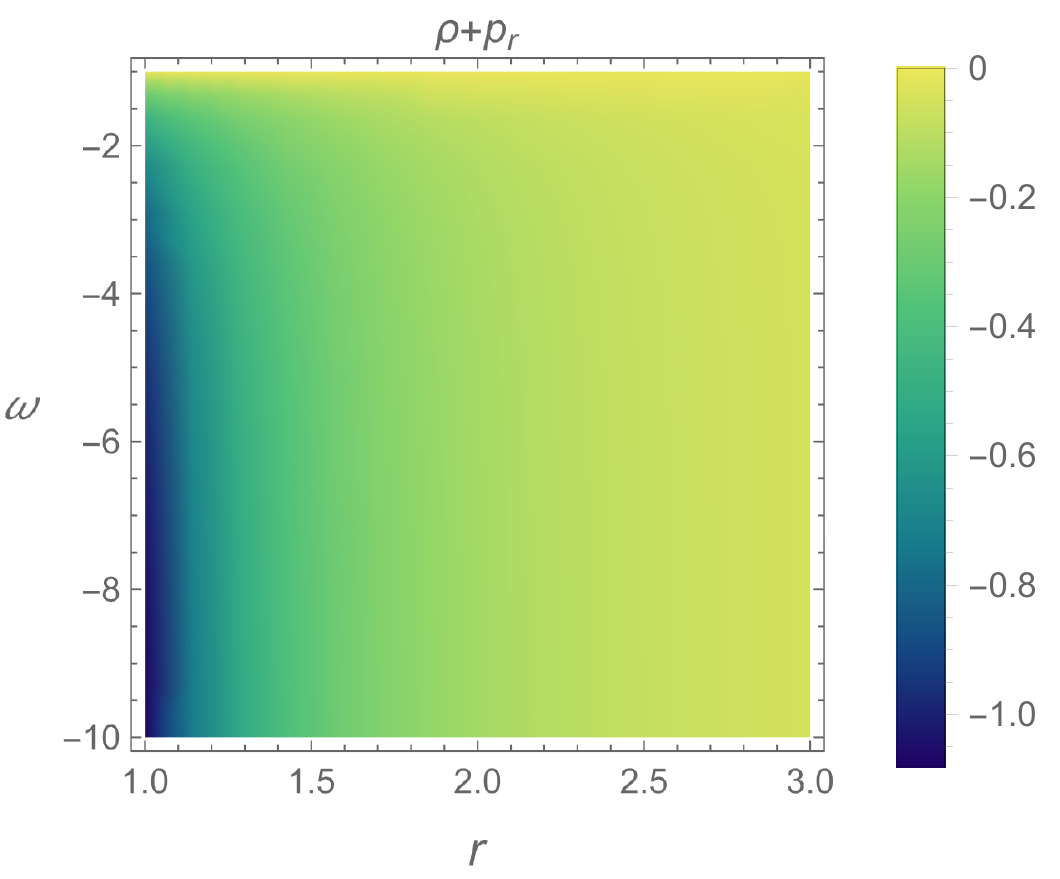}
	    \includegraphics[width=0.35\linewidth]{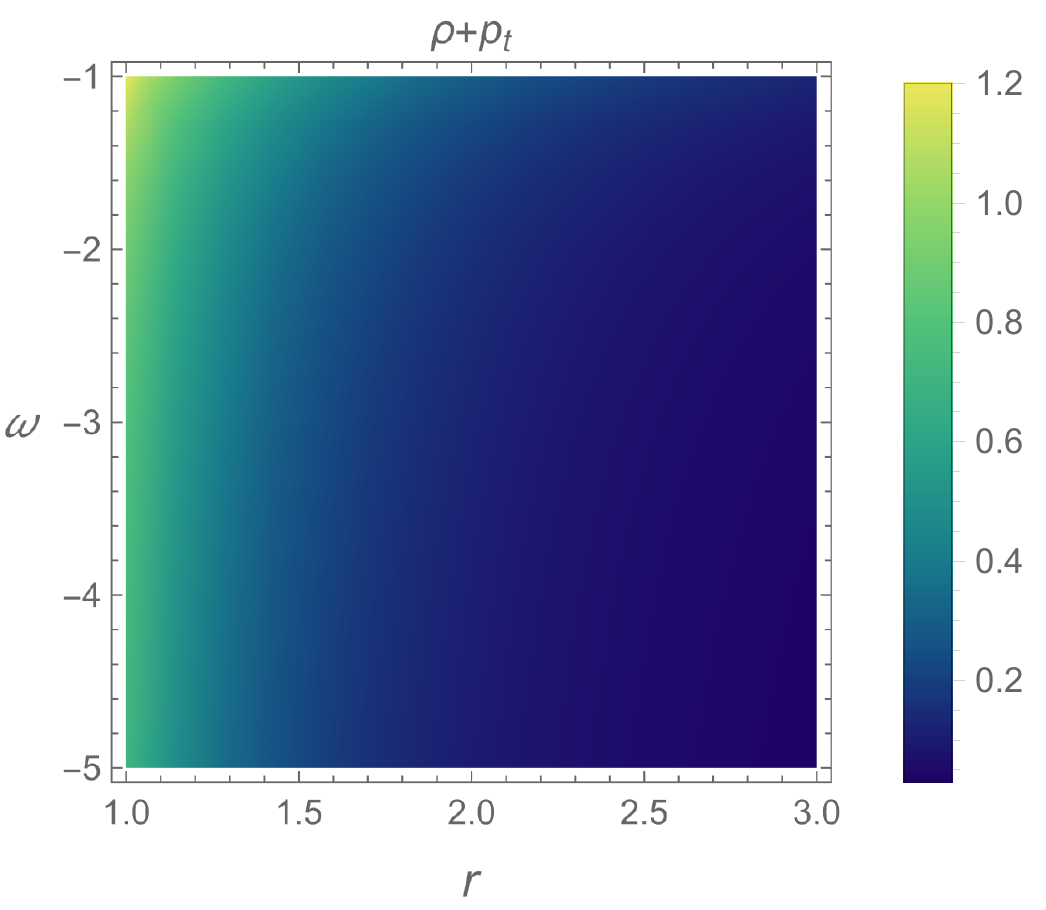}\\
	    \includegraphics[width=0.35\linewidth]{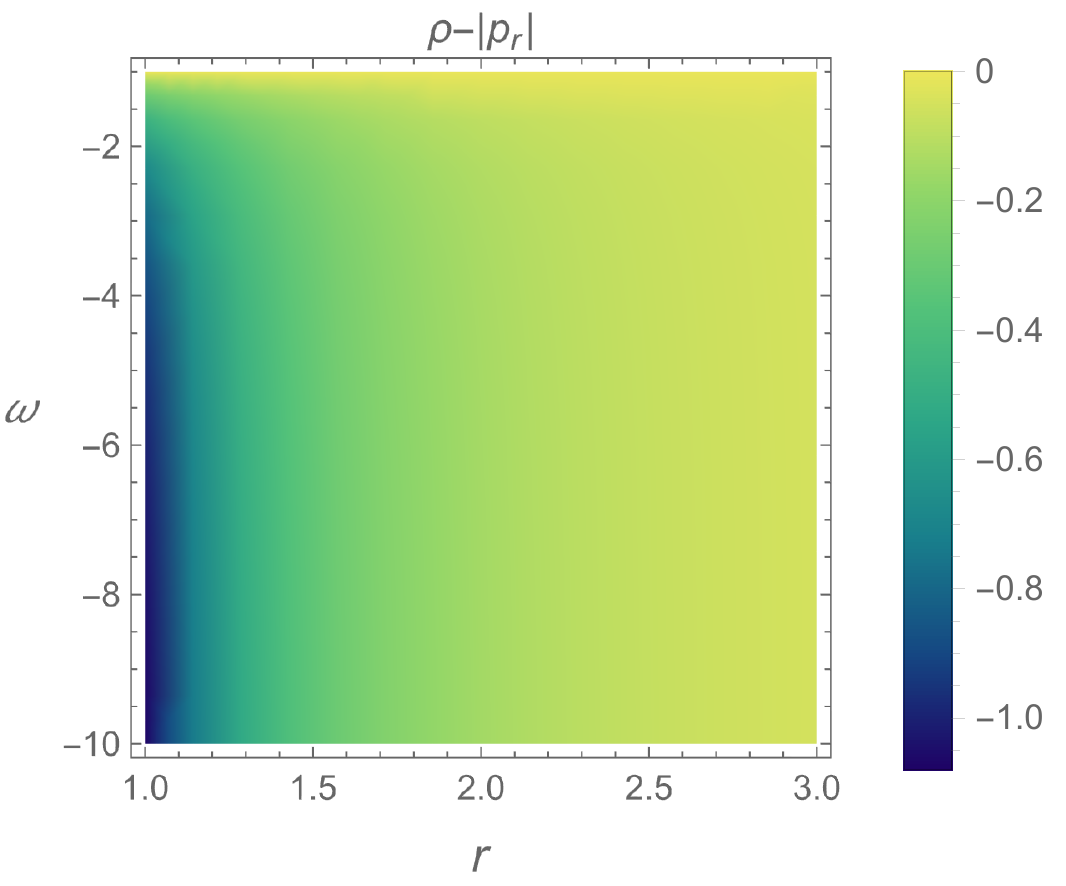}
	    \includegraphics[width=0.35\linewidth]{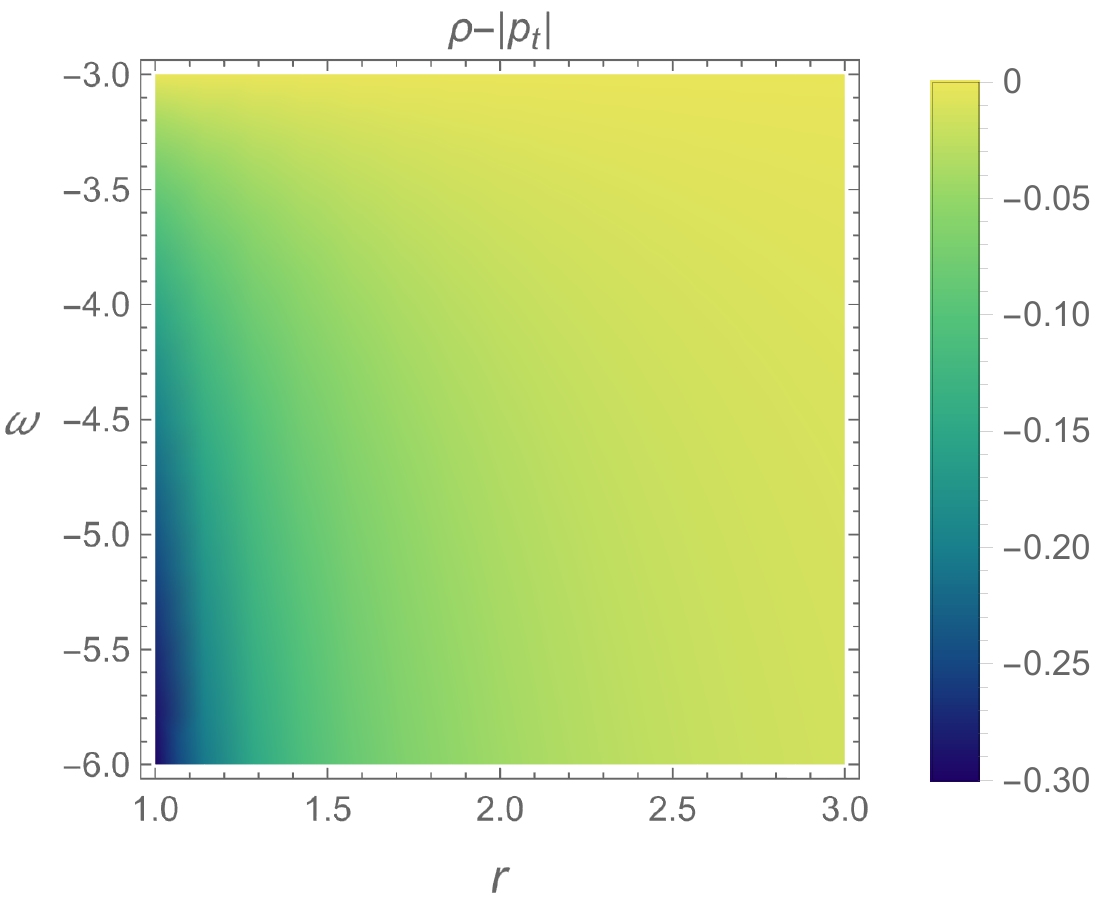}
	    \caption{WHM2: Profile of ECs with $\alpha=3, \beta=5, k_2=1$}
	    \label{fig:Bec}
	\end{figure*}

 It can be observed that, the energy density remains positive in the phantom region and in the range $(0,1]$ of the EoS parameter. Howerver, for the the quintessence $(-1<\omega<0)$, dust case $\omega=0$ and $\Lambda$CDM, the traverabality of WH is not satisfied. So, we restrict our domain of EoS parameter to $(-\infty,-1)\cup(0,1]$. This is summarized in \tableautorefname~ \ref{tab:table2}. Further, we can check the valid range for model parameters based on $\rho$. The behavior of different ECs is summarized in \tableautorefname~ \ref{tab:table3}. Also, for $\alpha=3, \beta=5, \omega =-2$ and $k_2=1$ the profile of ECs is given in \figureautorefname~ \ref{fig:Bec}.

		\begin{table}[h!]
		\caption{The overview of the result obtained based on the EoS parameter $\omega$.}
		    \label{tab:table2}
		\begin{ruledtabular}
		    \centering
		    \begin{tabular}{c c}
				 $\omega$ & Interpretations\\ 
				\hline
				$(-\infty,-1)$ & $\rho>0$ if $\alpha,\beta>0$ or $\alpha,\beta<0$ \\
				-1 & WH fails to be traversable \\
				$(-1,0)$ & WH fails to be traversable \\
				0 & WH fails to be traversable \\
				1/3 & $\rho>0$ if $\alpha>0,\beta<0$ or $\alpha<0,\beta>0$ \\
			\end{tabular}
		   \end{ruledtabular}
		\end{table}
	
	    \begin{table}[h!]
		\caption{The interpretation of energy conditions for case B with the set of values for $\omega$, $\alpha$, and $\beta$ for which $\rho>0$. In all intervals $\rho+p_r+2p_t=0$.}
		    \label{tab:table3}
		\begin{ruledtabular}
		    \centering
		    \begin{tabular}{c c c}
				 $\omega$ & $(-\infty,-1)$  & $(0,1]$\\ 
				\hline
				$\alpha$ and $\beta$  & $\alpha,\beta>0$ or  & $\alpha>0,\beta<0$ or  \\
				  &  $\alpha,\beta<0$  & $\alpha<0,\beta>0$ \\
				\hline
				$\rho$ & obeyed  & obeyed\\
				$\rho+p_r$ & violated & obeyed\\
				$\rho+p_t$ & obeyed  & violated\\
				$\rho-\mid p_r\mid$ & violated  & violated\\
				$\rho-\mid p_t\mid$ & violated  & violated\\
			\end{tabular}
		   \end{ruledtabular}
		\end{table}

\section{NON-LINEAR MODEL OF $\mathpzc{f}(\mathcal{R},\mathscr{L}_m)$ GRAVITY}\label{sectionVI}

     \subsection{Case: $\mathpzc{f}(\mathcal{R},\mathscr{L}_m)=\dfrac{1}{2}\mathcal{R}+\mathscr{L}_m^\eta$ (WHM3)}
	  \par We shall consider the non-linear model, 
	  \begin{equation}
	      \mathpzc{f}(\mathcal{R},\mathscr{L}_m)=\dfrac{1}{2}\mathcal{R}+\mathscr{L}_m^\eta,
	  \end{equation}
	  where $\eta$ is the model parameter. For $\eta=1$, one can obtain the GR scenario. The field equation can be written as, 
	  \begin{widetext}
	      \begin{align}
	          \label{eq1}\frac{b'}{r^2}+\eta  \rho^{\eta }-\rho^{\eta }=\eta  \rho^{\eta -1} (p_r+2 p_t+2 \rho),\\
	          \label{eq2}\frac{3 b-r b'}{r^3}-\frac{b'}{r^2}+\eta  \rho ^{\eta }-\rho^{\eta }=\eta  \rho ^{\eta -1} (-2 p_r+2 p_t-\rho),\\
	          \label{eq3}-\frac{3 b-r b'}{2 r^3}-\frac{b'}{r^2}+\eta  \rho ^{\eta }-\rho ^{\eta }=\eta  \rho ^{\eta -1} (p_r-p_t-\rho).
	      \end{align}
	  \end{widetext}
	  
	\par Now, solving \eqref{eq1} and \eqref{eq2} the radial and tangential pressure can be represented as,
	 
	 \begin{eqnarray}
	     \label{Pr}p_r=-\frac{\rho ^{1-\eta } \left(-r b'+b+\eta  r^3 \rho ^{\eta }\right)}{\eta  r^3},\\
	     \label{Pt}p_t=-\frac{\rho ^{1-\eta } \left(r^3 \rho^{\eta }-b\right)}{2 \eta  r^3}.
	 \end{eqnarray}

    \subsection*{Specific energy density: $\rho=\rho_0 \left(\frac{r_0}{r}\right)^m$ }

    Let us suppose a specific form of energy density,
	 \begin{equation}
	  \label{Rho} \rho=\rho_0 \left(\frac{r_0}{r}\right)^m.
	 \end{equation}
    Substituting the above energy density in \eqref{eq3}, and pressure elements \eqref{Pr}-\eqref{Pt} in \eqref{eq3}, we get,	
	  \begin{equation}\label{sf}
	      b(r)=k+\frac{(2 \eta -1) r^3 \left(\rho_0 \left(\frac{r_0}{r}\right)^m\right)^{\eta }}{3-\eta  m},
	  \end{equation}
    \par where $k$ is a constant of integration. In order to satisfy the throat condition, the shape function should obey the relation $b(r_0)=r_0$. Thus we have, 
    \begin{equation}
         k=\frac{r_0 \left(\eta  m-r_0^2 \rho_0^{\eta }+2 \eta  r_0^2     \rho_0^{\eta }-3\right)}{\eta  m-3}.
    \end{equation}
    Further, the flaring-out condition at the throat imposes the conditions on the choice of parameters $\eta$, $r_0$, and $\rho_0$. This can be expressed by the following inequality:
    \begin{equation}\label{inequality}
        (2 \eta -1) \rho_0^{\eta }<\frac{1}{r_0^2}.
    \end{equation}
     \par
\begin{figure}[h]
        \centering
        \includegraphics[width=0.8\linewidth]{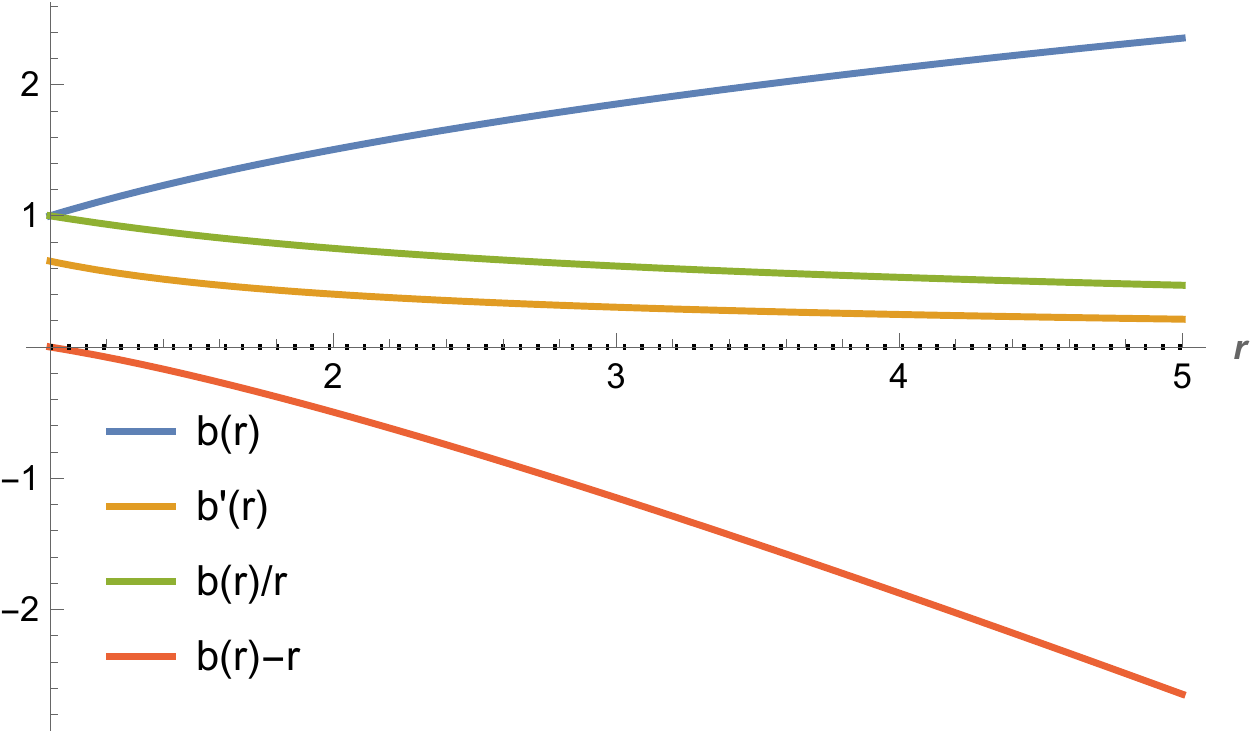}
        \caption{WHM3: Profile of shape function satisfying all the conditions for the values $\rho_0=0.8, \eta=0.9, m=3$ and $r_0=1$. }
        \label{fig:Dsf}
    \end{figure}

 The obtained shape function \eqref{sf} satisfies all the necessary conditions. For $\eta=0.9, r_0=1, \rho_0=0.8$ and $m=3$, the behaviour of shape function is represented in the \figureautorefname~$\;$\ref{fig:Dsf}. From \figureautorefname~$\;$\ref{fig:Dsf} it can be seen that the shape function $b(r)$ is a positive and monotonically increasing function. In the entire domain of $r$, $b(r)<r$ and $b(r)=r$ at the throat. The derivative of the shape function, $b'(r)$ tends to $0$ for $r\to\infty$ and $b'(r)<1$ for all $r$. Further, $b(r)/r\to0$ as $r\to\infty$.

    Using \eqref{Rho} and \eqref{sf}, equations \eqref{Pr} and \eqref{Pt} can be rewritten as,
    \begin{widetext}
        \begin{gather}
        p_r=\frac{1}{\eta  r^3 (\eta  m-3)}\left[\left(\rho_0 \left(\frac{r_0}{r}\right)^m\right)^{1-\eta } \left(r^3 (\eta  ((\eta -1) m-1)+2) \left(\rho_0 \left(\frac{r_0}{r}\right)^m\right)^{\eta }+r_0 (3-\eta  m)+(2 \eta -1) \left(-r_0^3\right) \rho_0^{\eta }\right)\right],\\
        p_t=-\frac{1}{2 \eta  r^3 (\eta  m-3)}\left[\left(\rho_0 \left(\frac{r_0}{r}\right)^m\right)^{1-\eta } \left(r^3 (\eta  (m+2)-4) \left(P \left(\frac{r_0}{r}\right)^n\right)^{\eta }+r_0 (3-\eta  m)+(2 \eta -1) \left(-r_0^3\right) \rho_0^{\eta }\right)\right].
    \end{gather}
    \end{widetext}
    \par
\begin{figure*}[t!]
	    \centering 
	    \includegraphics[width=0.3\linewidth]{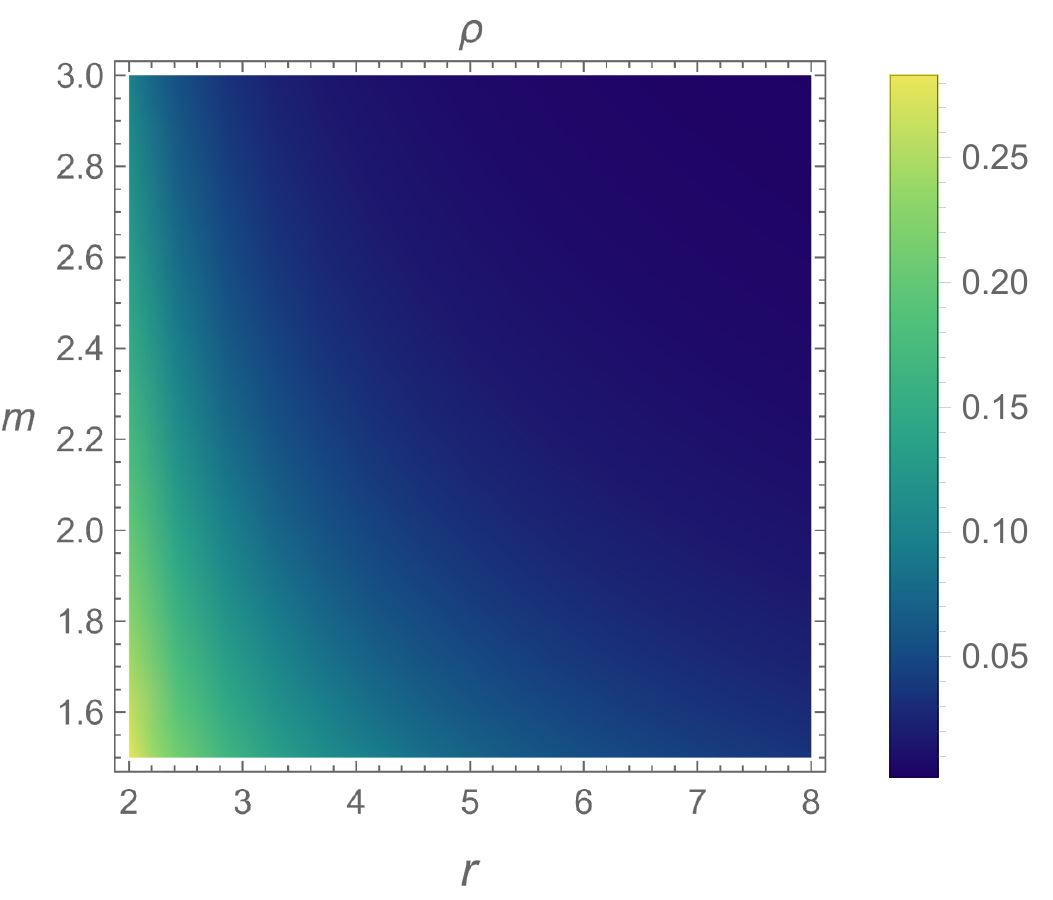}
	    \includegraphics[width=0.3\linewidth]{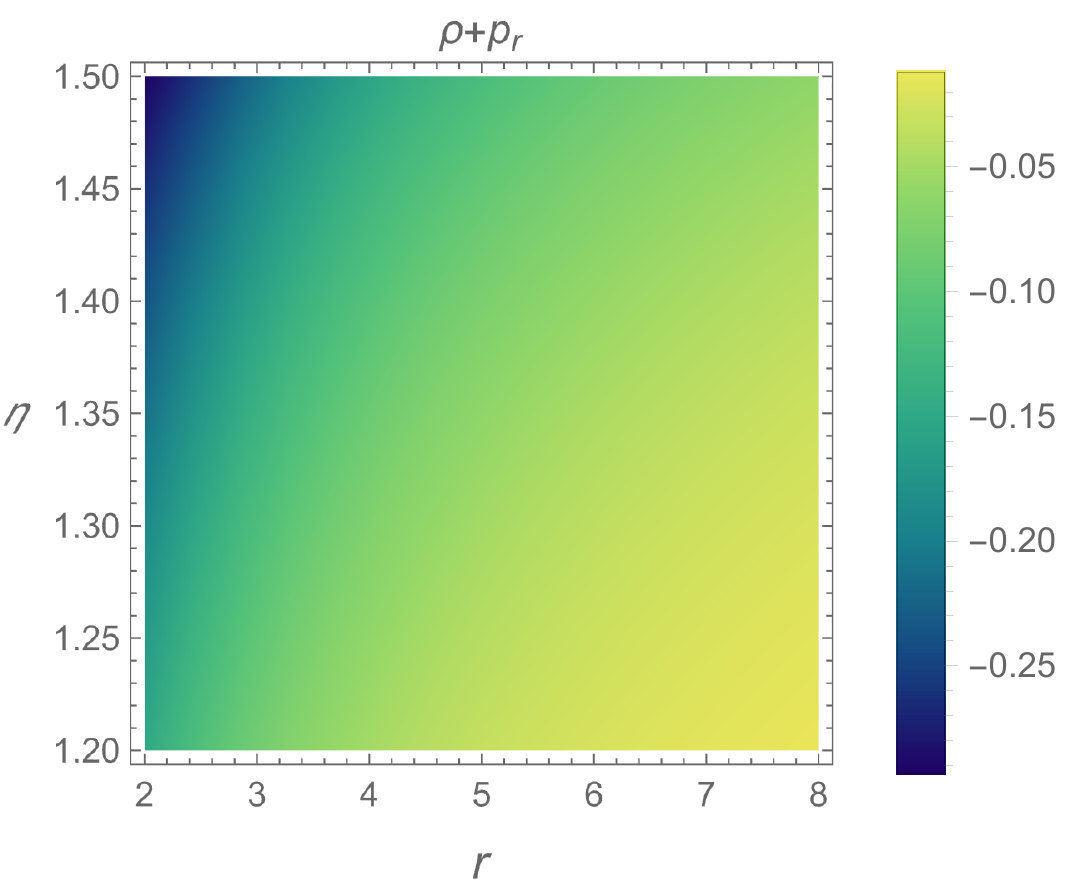}
	    \includegraphics[width=0.3\linewidth]{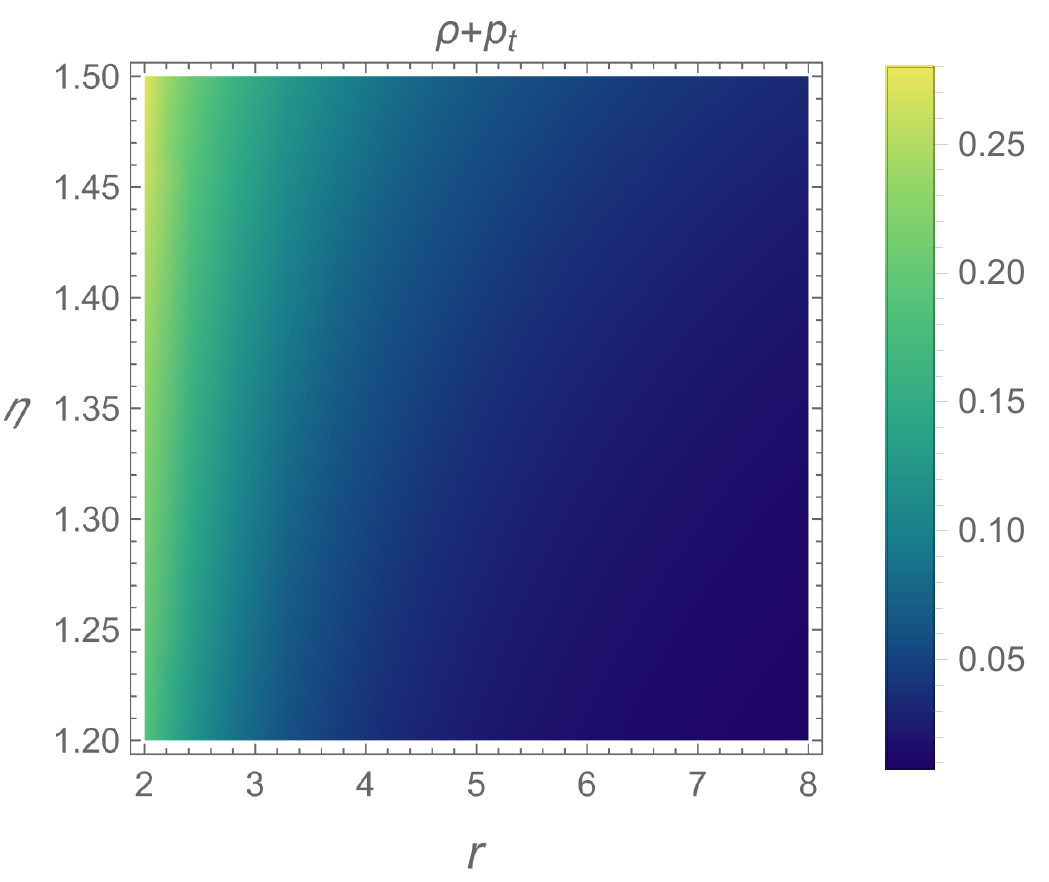}\\
	    \includegraphics[width=0.3\linewidth]{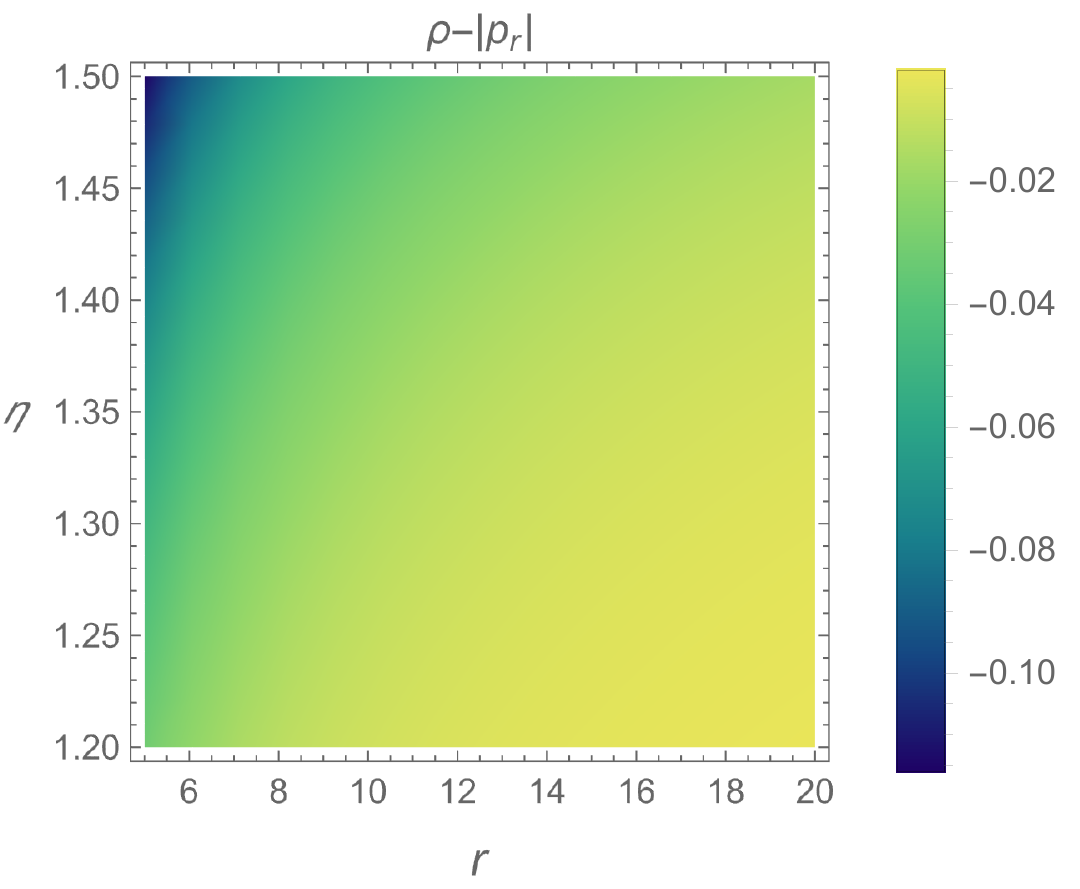}
            \includegraphics[width=0.3\linewidth]{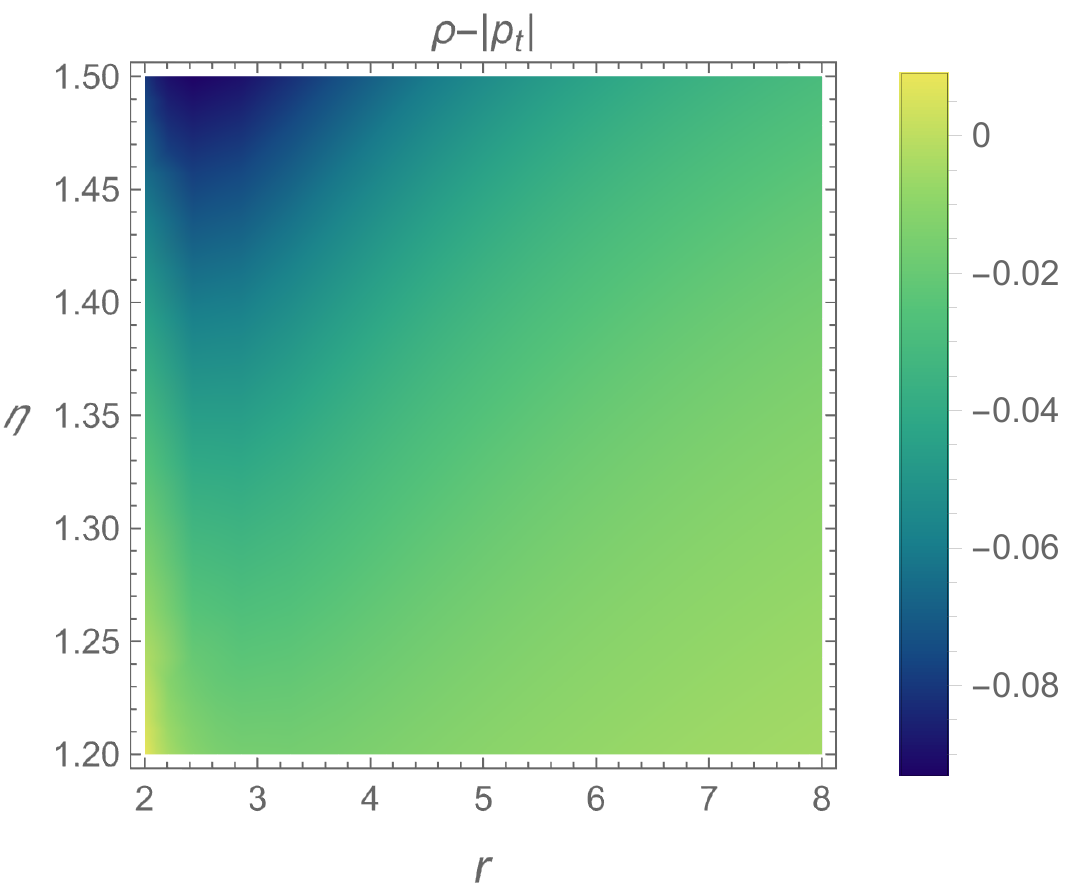}
            \includegraphics[width=0.3\linewidth]{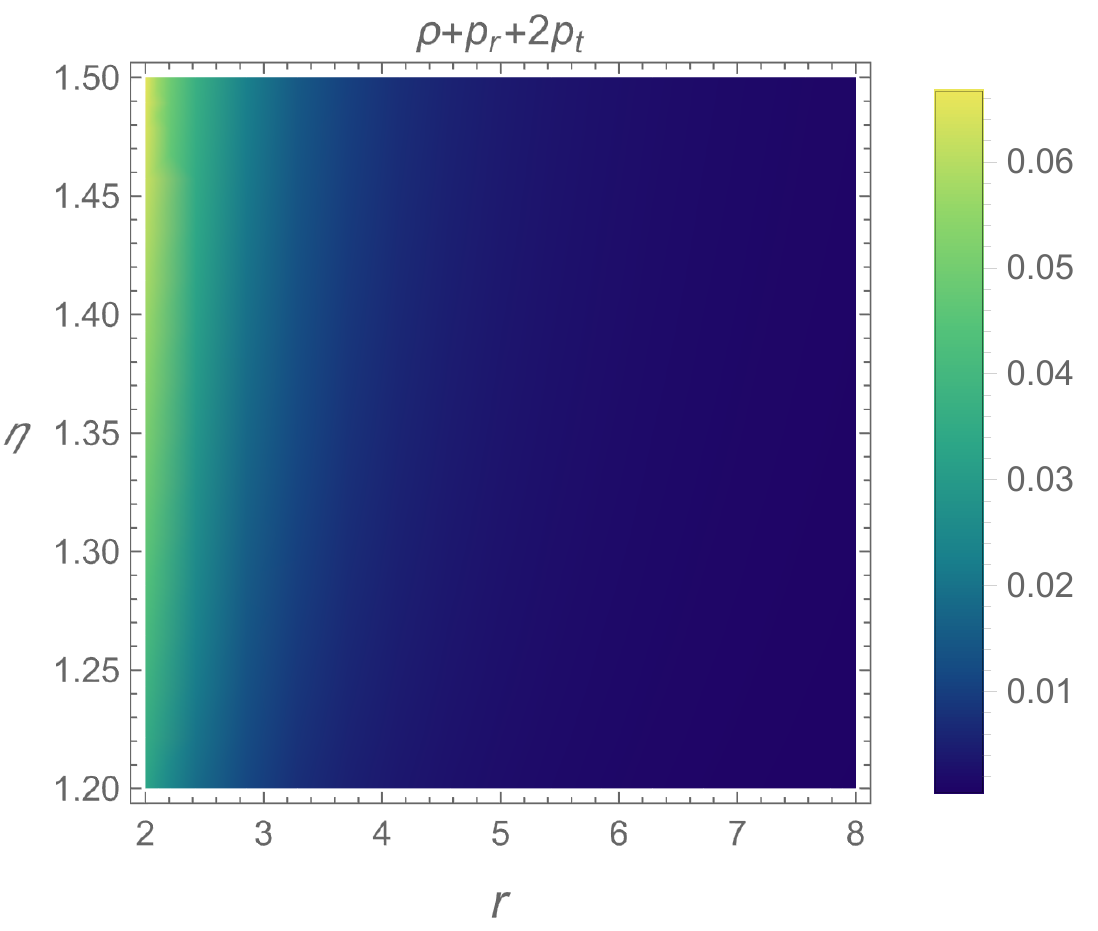}
	    \caption{WHM3: Energy density versus parameter $m$ (top-left panel) and profile of ECs varying w.r.t model parameter $\eta$ with $\rho_0=0.8, \eta=0.9, m=3$ and $r_0=1$.}
	    \label{fig:Dec}
    \end{figure*}

 With the help of the above expressions, we can analyze the energy conditions. The NEC for radial pressure is violated and for tangential pressure, it is obeyed. Both the DEC are violated and SEC is satisfied [\figureautorefname~$\;$\ref{fig:Dec}].

    \subsection{Case: $\mathpzc{f}(\mathcal{R},\mathscr{L}_m)=\dfrac{1}{2}\mathcal{R}+(1+\xi \mathcal{R})\mathscr{L}_m$ (WHM4)}
	In the present section, let us consider a non-linear model of $\mathpzc{f}(\mathcal{R},\mathscr{L}_m)$ gravity, defined by,
	\begin{equation}\label{nonlinear}
	    \mathpzc{f}(\mathcal{R},\mathscr{L}_m)=\dfrac{1}{2}\mathcal{R}+(1+\xi \mathcal{R})\mathscr{L}_m,
	\end{equation}
	where $\xi$ is a model parameter. When $\xi=0$, the study reduces to GR. Now for the current case, let us assume the matter distribution to be isotropic. Therefore, we can write, 
      $$p_r=p_t=p.$$

    \par Furthermore, the pressure element and the energy density are supposed to be related as,
    $$p=\omega \rho,$$ where $\omega$ is EoS parameter.
    
    With the above conditions the field equations \eqref{fe1}-\eqref{fe3} for the non-linear model \eqref{nonlinear} becomes,
    
    \begin{widetext}
        \begin{eqnarray}
            \label{nleq1}(4\xi\rho+1)b'=\rho(3\omega+2)(2\xi b'+r^{2}),\\
            \label{nleq2}b'-\left(2\xi\rho+1\right)\dfrac{(r b'-3b)}{2r}=6\xi(r-b)\rho'+(2\xi b'+r^{2})\rho,\\
            \label{nleq3}b'+\left(2\xi\rho+1\right)\dfrac{(r b'-3b)}{r}=6\xi(r-b)\rho''-6\xi r \rho'b'+(2\xi b'+r^{2})\rho.
        \end{eqnarray}
    \end{widetext}
	\subsection*{Specific Shape Function: $b(r)=r_0 \left(\dfrac{r_0}{r}\right)^n$ }
	
	\begin{figure}[h!]
		    \centering
		    \includegraphics[width=0.8\linewidth]{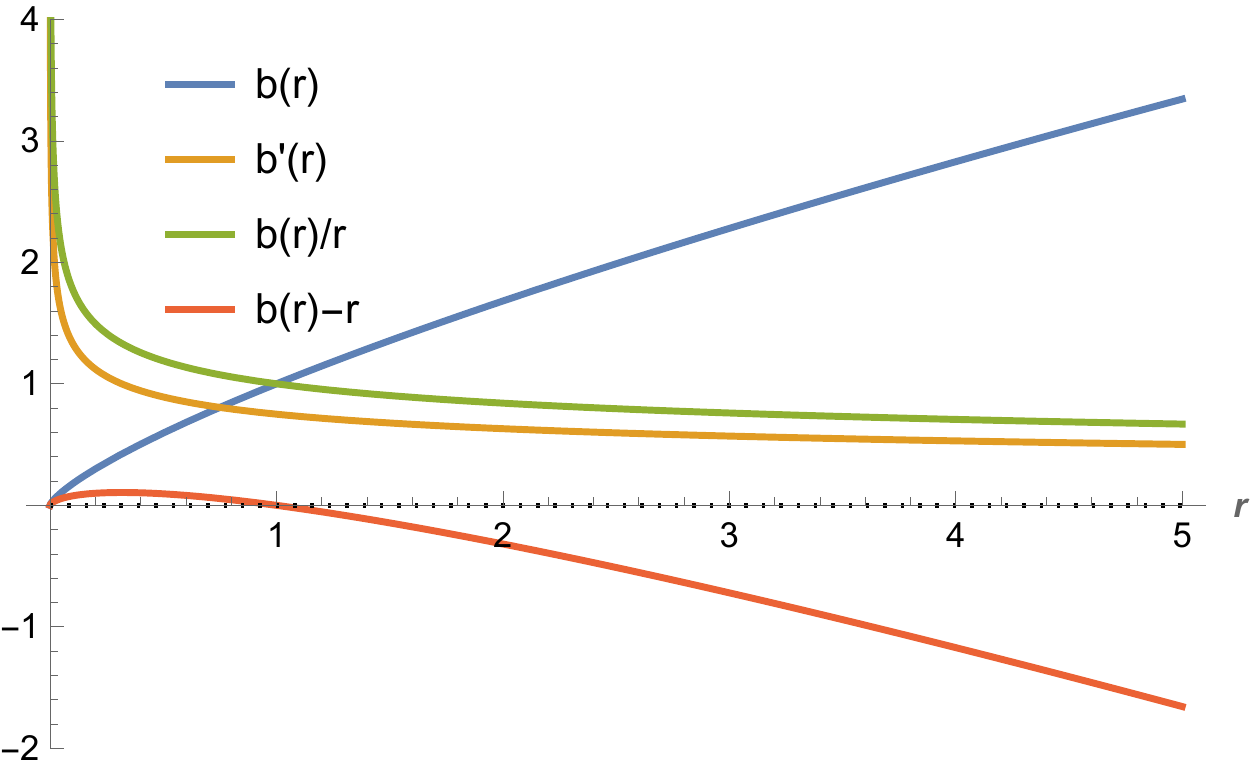}
		    \caption{WHM4: Profile of the shape function $b(r)=r_0 \left(\frac{r_0}{r}\right)^n$ with $n=-0.75$, $r_0=1$ and satisfying $\frac{b(r)}{r}<1$, $b'(r)<1$ and $\frac{b(r)}{r}\rightarrow 0$ as $r\rightarrow \infty$}
		    \label{fig:sfC}
		\end{figure}
	\par Now, let us consider a specific shape function, 
	\begin{figure*}[!]
	    \centering
	    \includegraphics[width=0.35\linewidth]{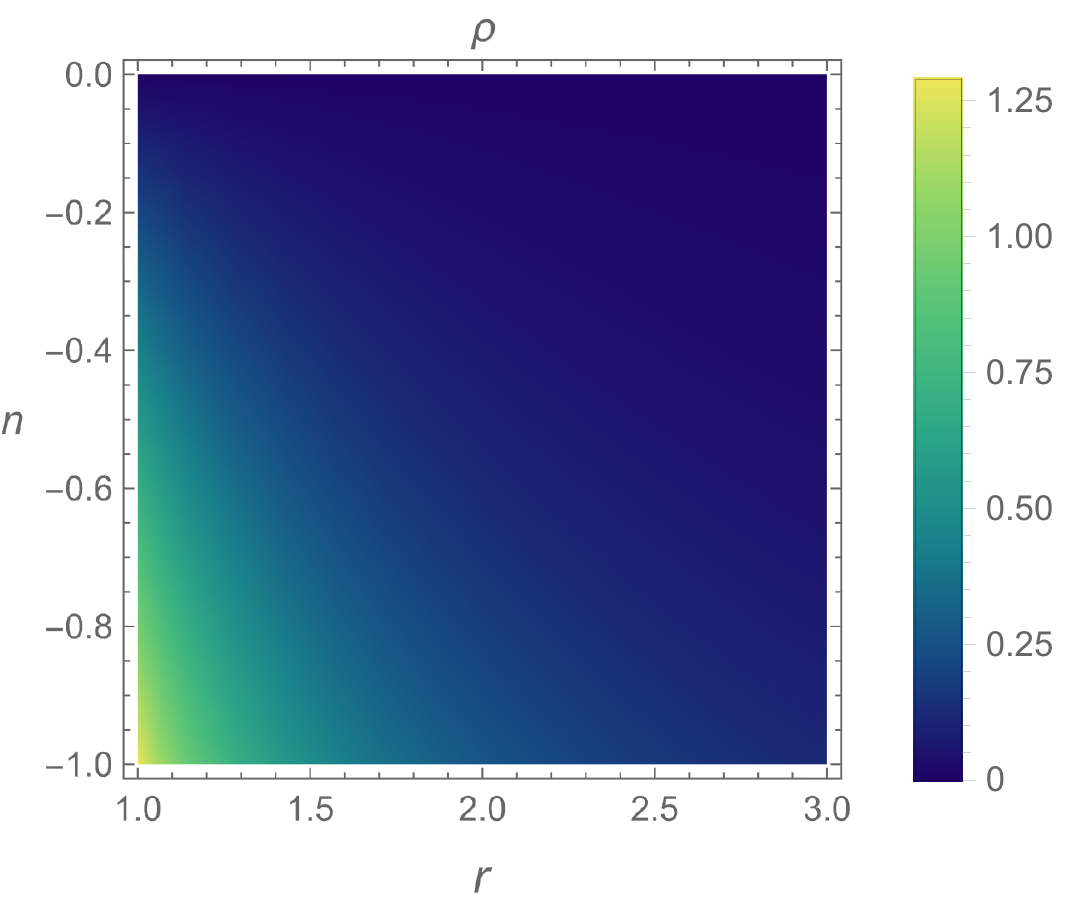}
	    \includegraphics[width=0.35\linewidth]{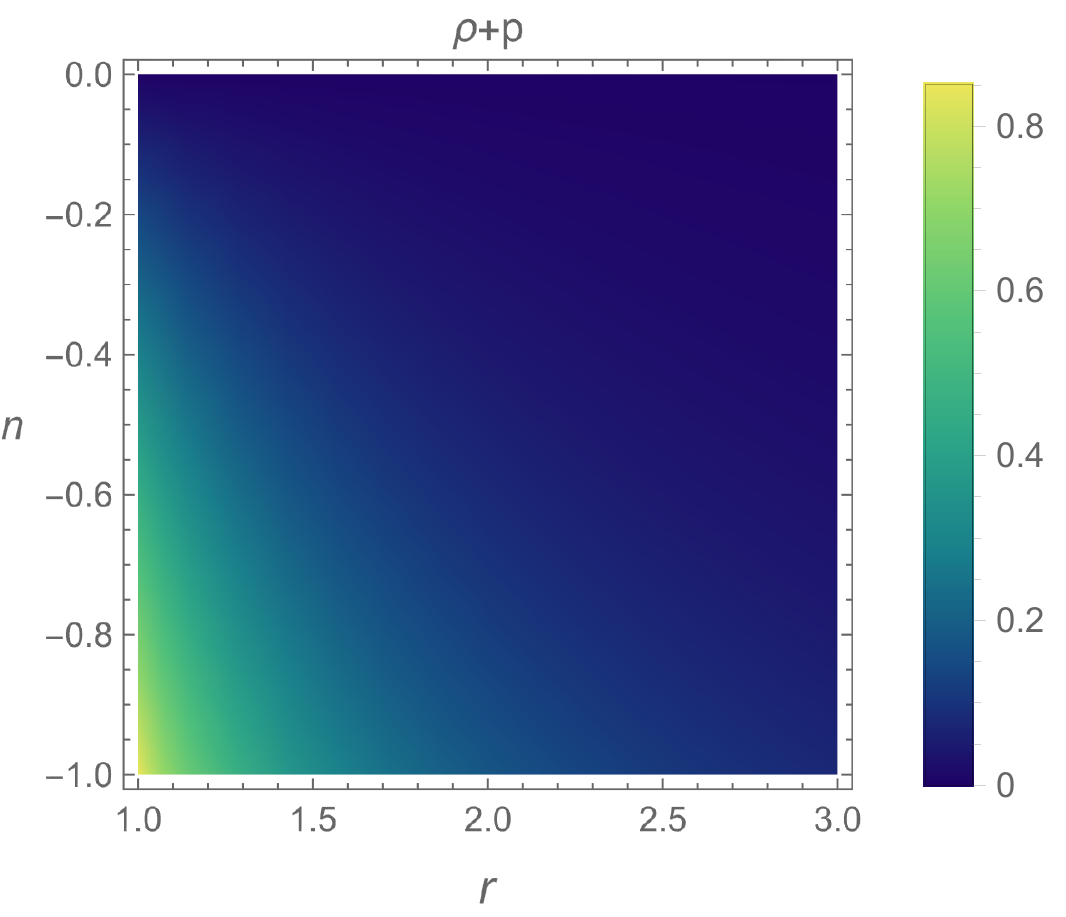}\\
	    \includegraphics[width=0.35\linewidth]{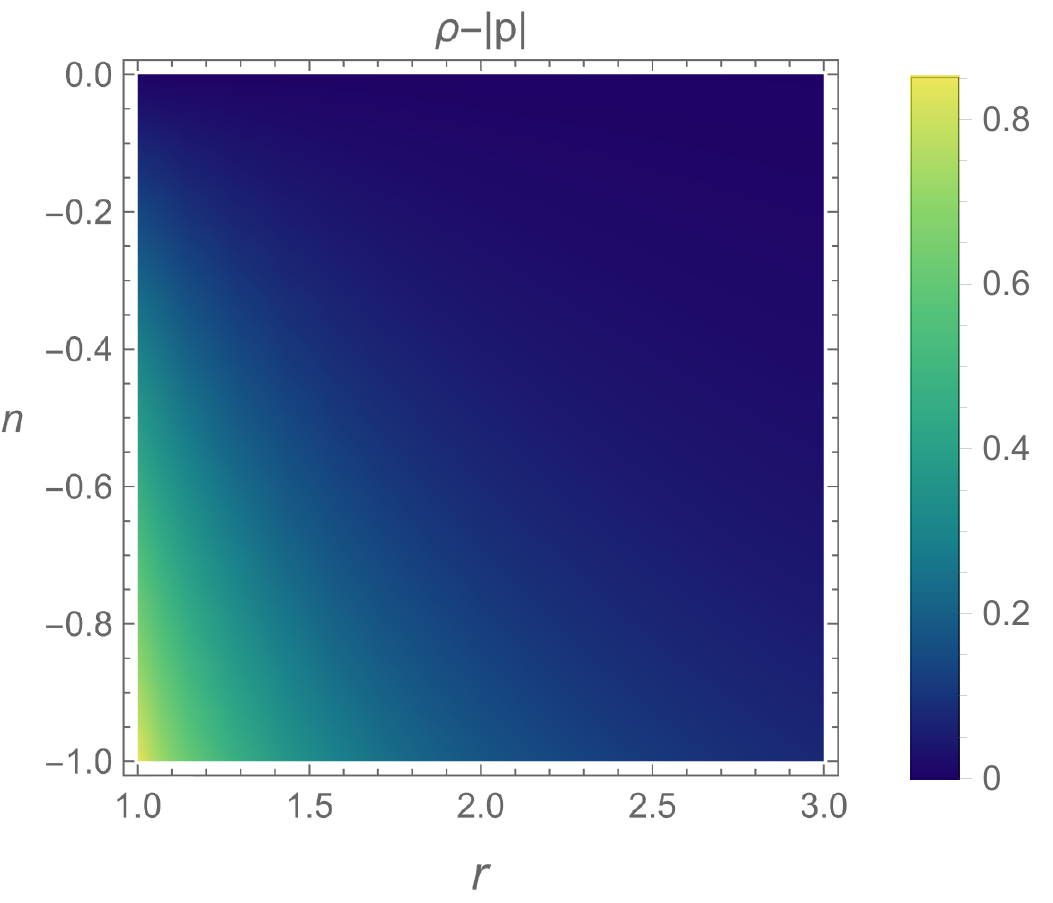}
	    \includegraphics[width=0.35\linewidth]{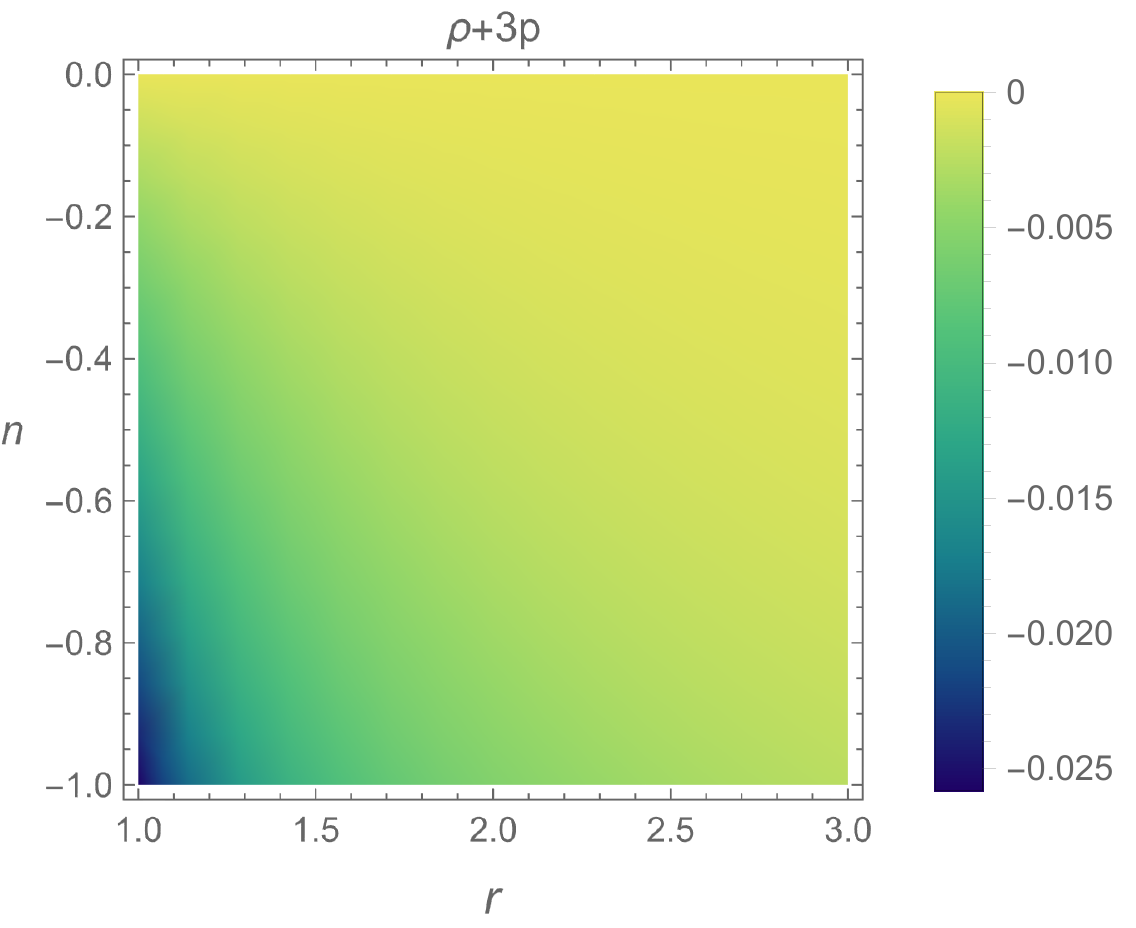}
	    \caption{WHM4: Energy Condition Profile for Non-linear case with $\xi=0.1$ and $\omega=-0.34$}
	    \label{fig:Cec}
	\end{figure*}

	\begin{equation}\label{specialsf}
	    b(r)=r_0 \left(\dfrac{r_0}{r}\right)^n,
	\end{equation} where $n$ is an arbitrary real value. In order to satisfy the traversability condition for a WH, $n$ should lie in $(-1,0)\cup(0,\infty)$. This can be seen in \figureautorefname~ \ref{fig:sfC} where we illustrated for values $n=-0.75$ and $r_0=1$. Now with the equation \eqref{specialsf}, taking $r_0=1$ the equation  \eqref{nleq1} yields,
	\begin{equation}
	    \rho=\dfrac{n(\frac{1}{r})^{n}}{6(\frac{1}{r})^{n}\xi n\omega-r^{3}(3\omega +2)}.
	\end{equation}
	\par To make this energy density positive in the whole spacetime the range of $n$ is constrained to $(-1,0)$. Further, in the quintessence region, particularly in $ (-2/3,-1/3)$, density remains positive. Also, the range of values for the model parameter $\xi$ is taken to be $[-0.1,0.1]$. It is to be noted that, one can take any other value of $\xi$, but for the current analysis we consider the mentioned range. In addition, the \figureautorefname~ \ref{fig:Cec} shows that NEC is satisfied. This indicates the absence of exotic matter at the WH throat. Also, DEC is obeyed. But there is a violation of the SEC.

\section{Embedding Diagram}\label{sectionVII}
		
		\par The embedding diagrams are very advantageous in depicting the visualized insights of WH. It depends on the choice of the shape function $b(r)$. As we are using a spherically symmetric metric for the present WH structure, we shall focus on the equatorial slice given by, $\theta=\frac{\pi}{2}$. Further, the value of the time coordinate is fixed, so that $t= constant$. On applying these conditions to \eqref{whmetric}, the metric becomes,
		\begin{equation}\label{eq:metricembedding1}
			ds^2 = \dfrac{dr^2}{1-\frac{b(r)}{r}}+r^2 \; d\phi^2.
		\end{equation}  

		\par Now, one can embed the above slice into its hypersurface with $(r,\phi,z)$, where $r,\phi\text{ and }z$ represents cylindrical coordinates. This metric is given by,	
		
		\begin{equation}\label{eq:metricembedding2}
			ds^2 = dz^2+dr^2+r^2 \; d\phi^2.
		\end{equation}

		\par From \eqref{eq:metricembedding1} and \eqref{eq:metricembedding2} we can find an expression for $z(r)$ that reads,
		\begin{equation}
			\dfrac{dz}{dr}=\pm\left[ \dfrac{b(r)}{r-b(r)}\right] ^{1/2}.
		\end{equation}
		
		\par
\begin{figure}[!]
		    \centering
		     \includegraphics[width=0.9\linewidth]{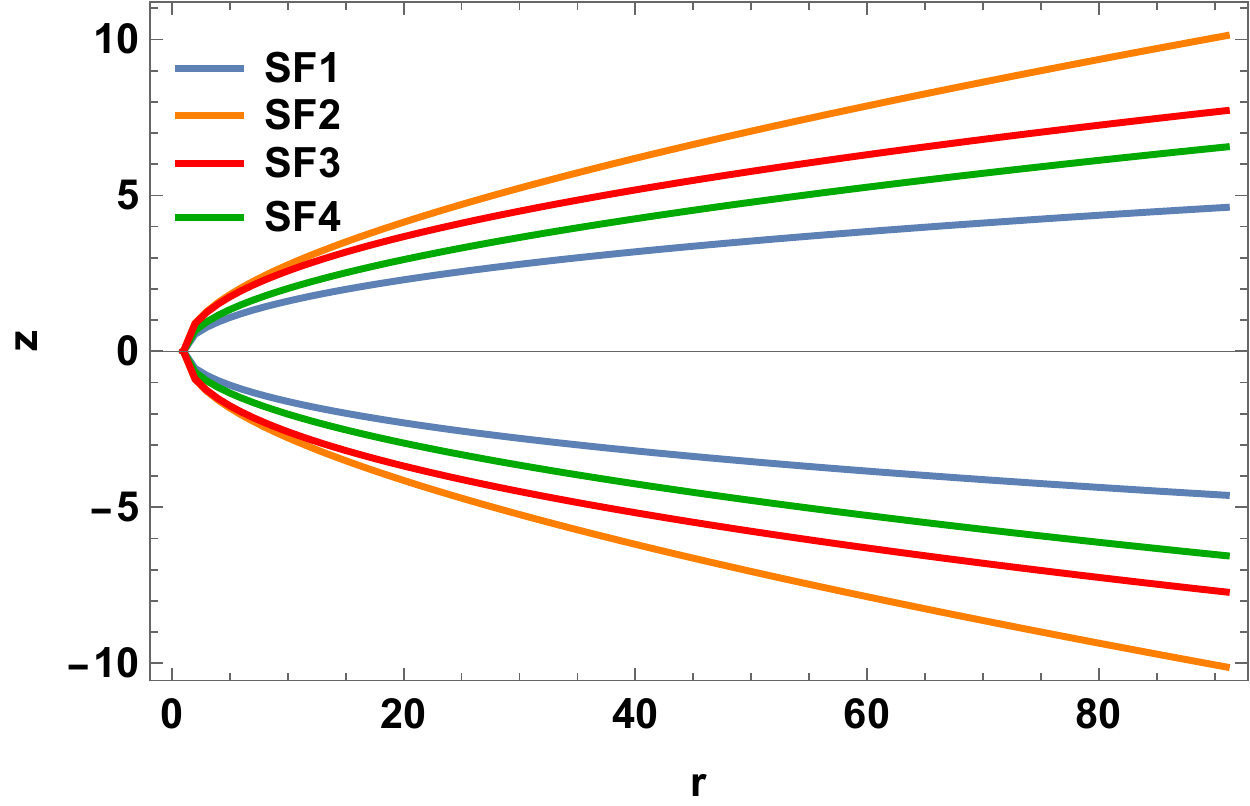}
		    \caption{Two dimensional embedding diagrams for WHM1 with $m=-3$ and $k_1=1$, WHM2 with $\omega=-2$ and $k_2=1$, WHM3 with $\rho_0=1, \eta=1, m=4$ and $r_0=1$ and WHM4 having the specific shape function $b(r)=r_0 \left(\dfrac{r_0}{r}\right)^n$ with $r_0=1$ and $n=-0.1$}
		    \label{fig:Ed2}
		\end{figure}

 The two-dimensional and three-dimensional embedding diagrams for different WH models are represented in \figureautorefname~ \ref{fig:Ed2}.  
			\begin{figure}[!]
		    \centering
		    \includegraphics[width=0.7\linewidth]{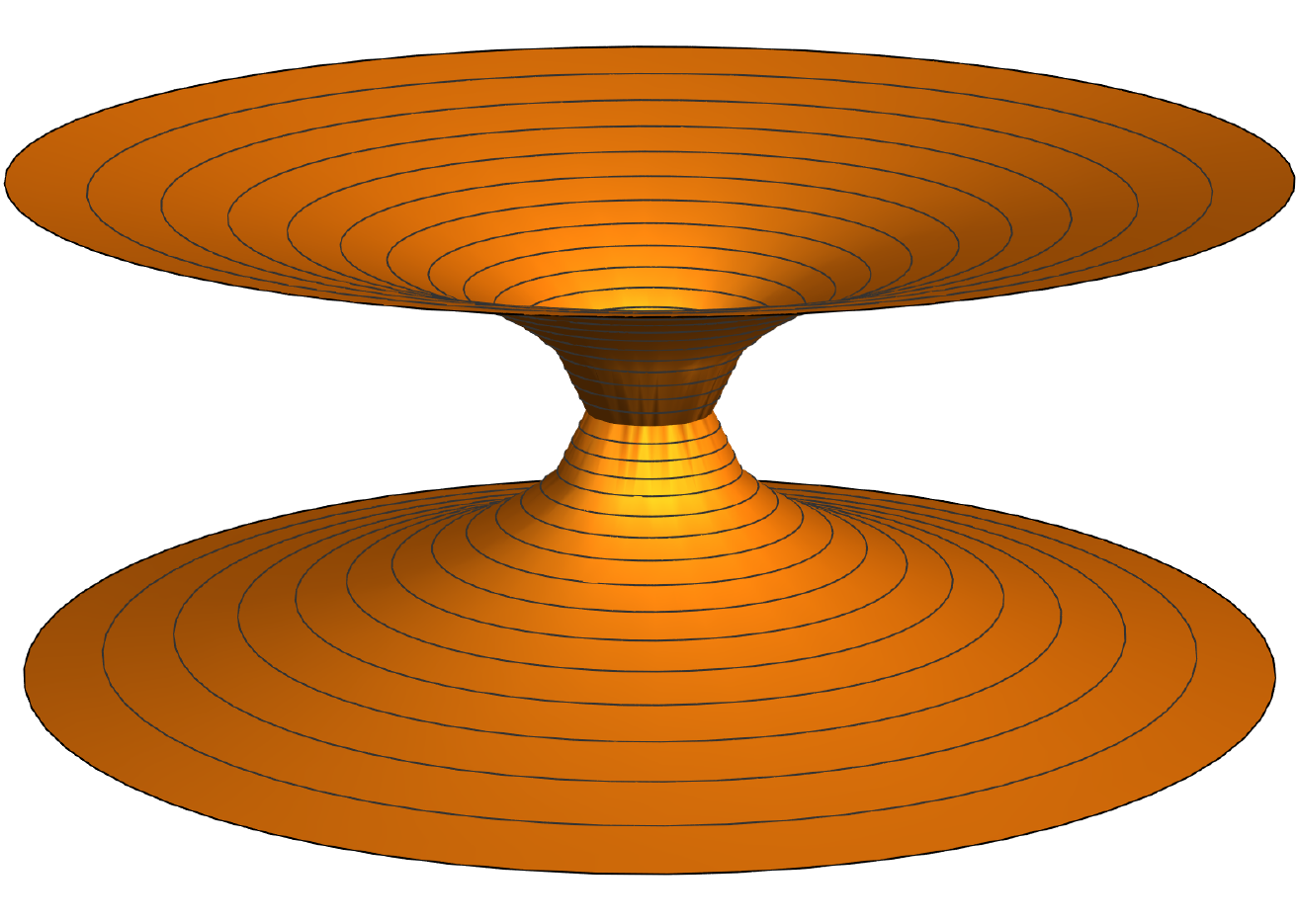}{a}
		    \includegraphics[width=0.7\linewidth]{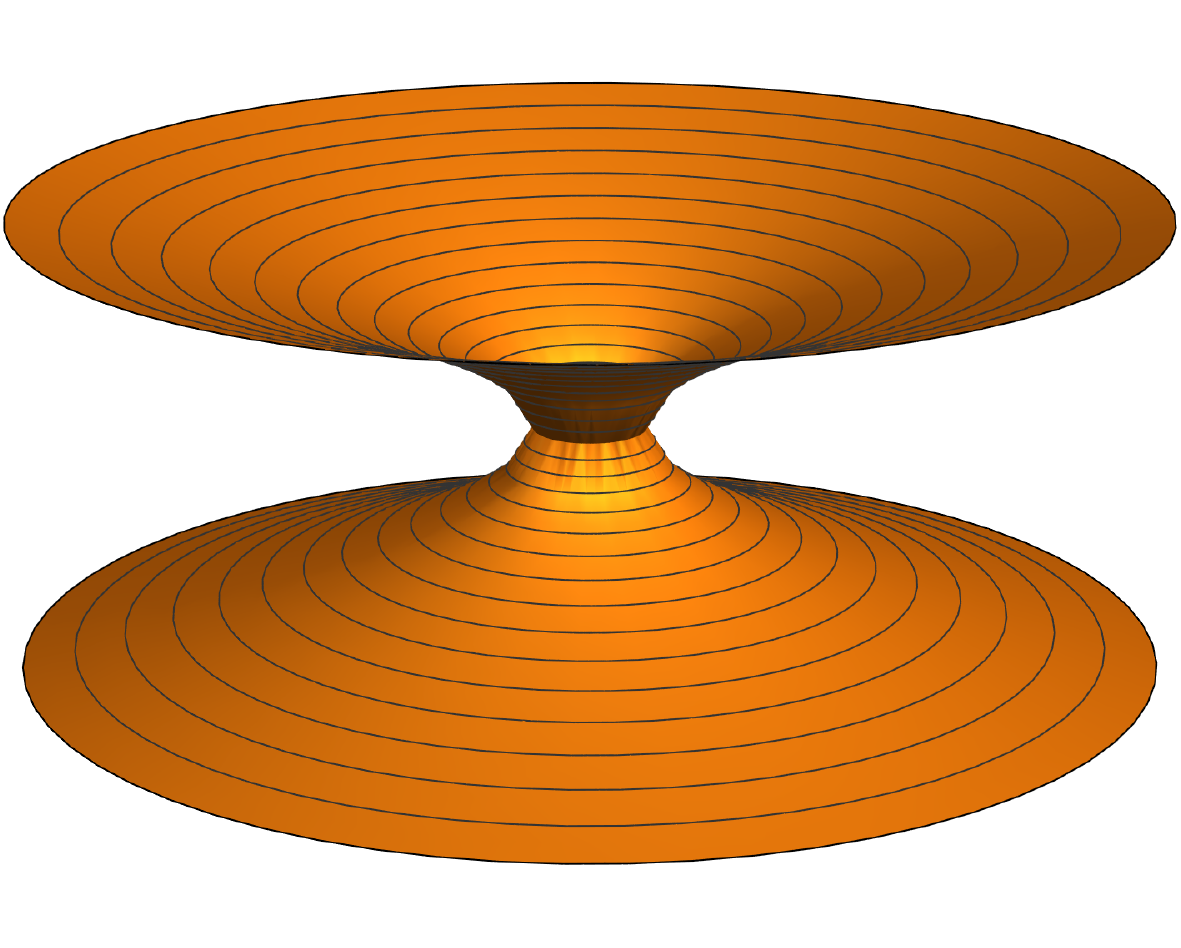}{b}
		    \includegraphics[width=0.7\linewidth]{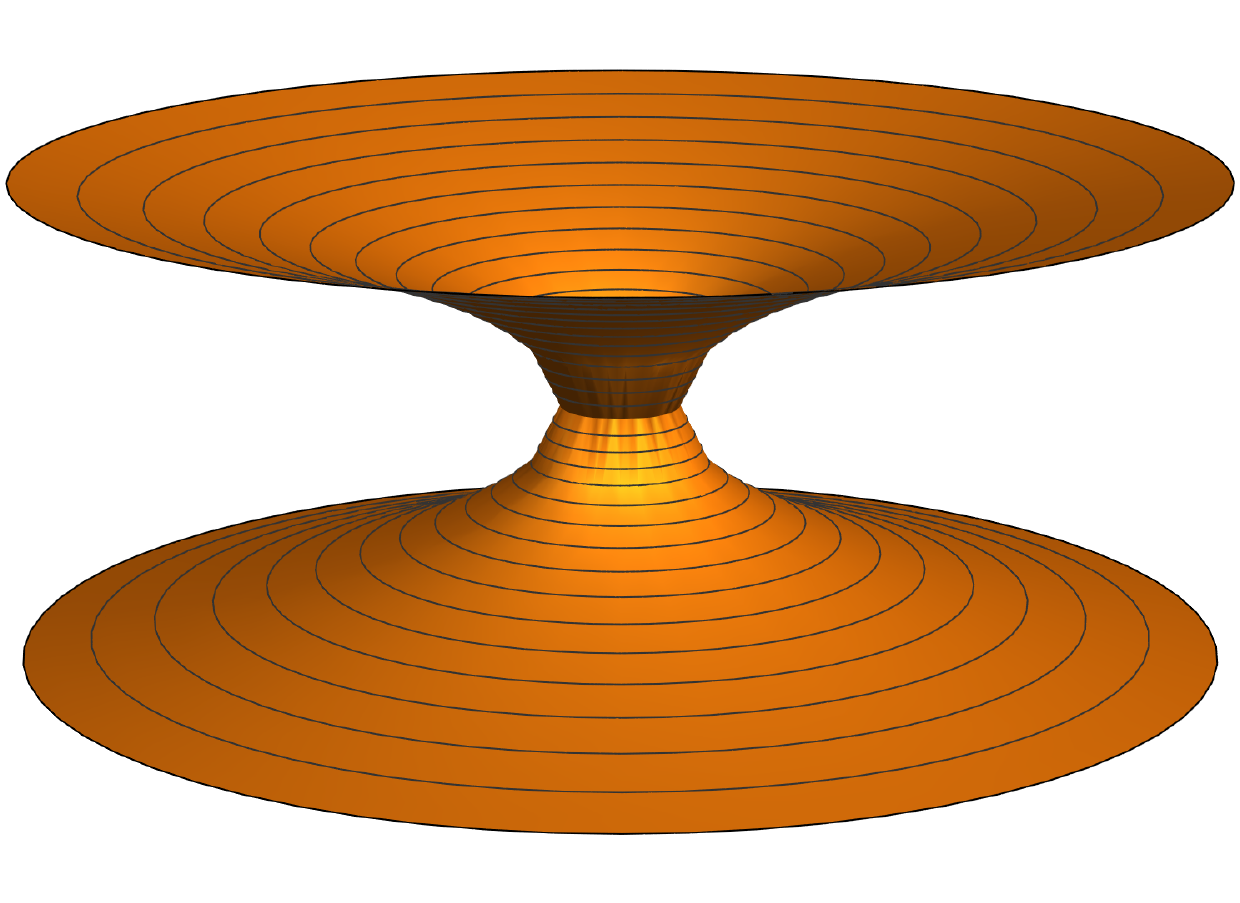}{c}
		    \includegraphics[width=0.7\linewidth]{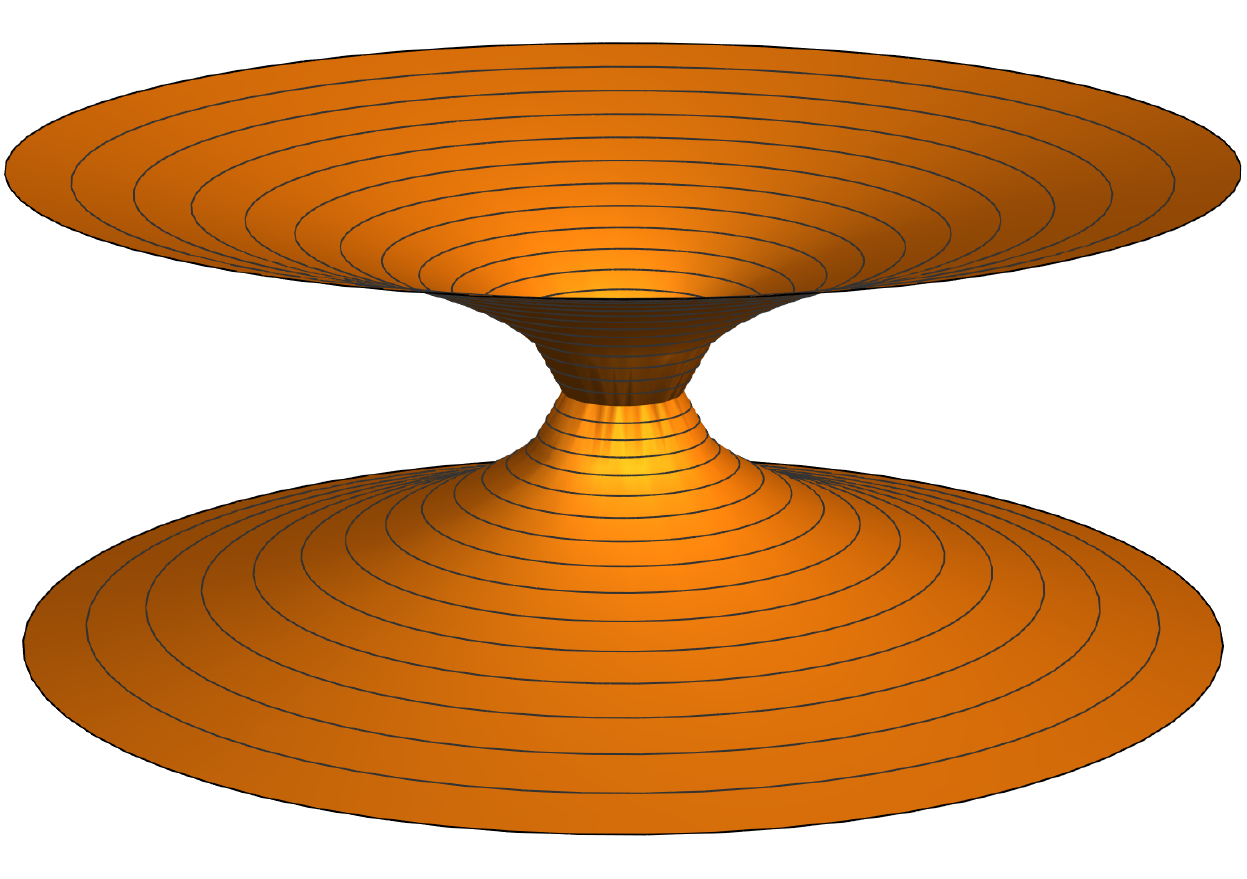}{d}
		    \caption{Three dimensional embedding diagrams for (a) WHM1 with $m=-3$ and $k_1=1$, (b) WHM2 with $\omega=-2$ and $k_2=1$, (c) WHM3 with $\rho_0=1, \eta=1, m=4$ and $r_0=1$ (d)  WHM3 having the specific shape function $b(r)=r_0 \left(\dfrac{r_0}{r}\right)^n$ with $r_0=1$ and $n=-0.1$ }
		    \label{fig:Ed3}
		\end{figure}
		
	
\section{Discussion and Final Remarks}\label{sectionVIII}
		\par In recent years, WH with its special geometric features, has evolved as one of the interesting topics of study in modern cosmology. Right from Flamm's solution to the present humanly traversable WH, many endeavors have been taken up to bring this mathematically drawn solution to support the physical reality. For instance, the unstable solution was improvised by a bridge-like structure, and the ordinary WH was devised as a traversable WH.  All these investigations were made in the background of GR. However, the physical existence of such a structure was put unanswered as there required a hypothetical fluid called exotic matter for the existence of a traversable WH. Later, it was found that modified theories can repudiate the existence of this exotic matter. In dealing with such a scenario, modified theories with matter couplings work reasonably well. In this regard, in the present work, we examined a spherically symmetric Morris-Thorne traversable WHs within the framework of $\mathpzc{f}(\mathcal{R},\mathscr{L}_m)$ gravity.

        \begin{itemize}
            \item For the WH to be traversable the redshift function should have a finite value everywhere in the domain. Therefore, in our work, we considered it as a constant. In addition, the matter distribution is assumed to be anisotropic. 
            
            \item Firstly, a linear model of $\mathpzc{f}(\mathcal{R},\mathscr{L}_m)$ gravity, $\mathpzc{f}(\mathcal{R},\mathscr{L}_m)=\alpha \mathcal{R}+\beta \mathscr{L}_m$, which can serve as GR equivalent is studied. The investigation of WH is done in two scenarios depending on the linear EoS conditions. Here, we examined the WH solutions for the case, $p_r = m p_t$ (linearly related radial and tangential pressures). Then, based on the EoS parameter $\omega$ the solution is analyzed. The shape functions obtained in both circumstances were found to obey the fundamental criteria for the traversable WH. The interpretation is done for different values of $\omega$ such as phantom, $\Lambda$CDM, quintessence, dust, and matter. For both WHs one can verify the violation of NEC, WEC, and DEC. Also, $\rho+p_r+2p_t=0$ (\tableautorefname~ \ref{tab:table1}, \ref{tab:table3}). 

            \item Secondly, we analyzed a minimal non-linear form of $\mathpzc{f}(\mathcal{R},\mathscr{L}_m)$ gravity,  $\mathpzc{f}(\mathcal{R},\mathscr{L}_m)=\dfrac{1}{2}\mathcal{R}+\mathscr{L}_m^\eta $. Here, we considered a specific form of energy density $\rho=\rho_0 \left(\frac{r_0}{r}\right)^m$ and we derived an expression for the shape function. For the derived shape function, the constant of integration is determined using the throat condition ($b(r_0)=r_0$) as the initial condition. Further, inequality \eqref{inequality} is so remarkable. It describes the constraining relation between free parameters. For the GR case, $\eta=1$. Then, \eqref{inequality} reads, $\rho_0r_0^2<1$. This relation predicts the valid parameter space for the energy density at the WH throat for a given throat radius i.e. $0\le\rho_0<1/r_0^2$. The obtained shape function $b(r)$ satisfies throat, flaring-out and asymptotic flatness conditions [\figureautorefname~~ \ref{fig:Dsf}]. 

            \item  Next, we investigated a non-minimal non-linear forms of $\mathpzc{f}(\mathcal{R},\mathscr{L}_m)$ gravity, $\mathpzc{f}(\mathcal{R},\mathscr{L}_m)=\dfrac{1}{2}\mathcal{R}+(1+\xi \mathcal{R})\mathscr{L}_m$. Field equations obtained for the non-minimal $\mathpzc{f}(\mathcal{R},\mathscr{L}_m)$ coupling are highly non-linear. Due to the complexities in obtaining an analytical solution, we presumed the isotropic matter distribution ($p_r=p_t=p$) with EoS relation $(p=\omega \rho)$ and the specific power-law form of shape function $b(r)=r_0 \left(\dfrac{r_0}{r}\right)^n$. This type of scenario is considered by Cataldo et al in \cite{cataldo}. 

            \item In both non-linear models, we studied the influence of a small amount of matter coupling. In model \eqref{nonlinear}, we considered the coupling constant $\xi$ in the range $[-0.1,0.1]$. In comparison with the linear model, we can interpret that a small amount of matter coupling can influence the geometry of WH.

            \item In the first gravity model that mimics GR, the NEC is violated which represents the presence of exotic matter at the WH throat. However, in the non-minimal model (WHM4), NEC is satisfied implying the absence of hypothetical fluid (one can refer \cite{1ref1,1ec1,1ec2,1ec3} for similar results). For WHM3, NEC for tangential pressure is satisfied, but for radial pressure it is violated. This indicates the necessity of exotic matter. The SEC is satisfied for the minimal model and is violated for the non-minimal model. For the linear model, it is 0 in the entire domain.

            \item  In our entire study, we considered parameter space for our model parameter for which energy density is positive. Further, all plots are plotted for the valid choice of these parameters. Lastly, we plotted 2-dimensional and 3-dimensional embedding diagrams for WH models.
        \end{itemize}
        
		\par To conclude, we investigated traversable WH in $\mathpzc{f}(\mathcal{R},\mathscr{L}_m)$ gravity, and interestingly our model upholds the absence of exotic fluid for non-minimal gravity model. In near future, we can examine some more observational interpretations of the WH solution with $\mathpzc{f}(\mathcal{R},\mathscr{L}_m)$ gravity models.
		
\section*{Data Availability Statement}
There are no new data associated with this article.

\begin{acknowledgments}
 V.V. and N.S.K. acknowledge DST, New Delhi, India, for its financial support for research facilities under DST-FIST-2019.

\end{acknowledgments}

	
\end{document}